\documentclass[aps,prb,twocolumn,superscriptaddress,longbibliography, nofootinbib]{revtex4-2}

\usepackage[utf8]{inputenc}
\usepackage{lineno}
\usepackage{hyperref}
\usepackage{physics}
\usepackage{nccmath}
\usepackage{empheq}
\usepackage{amsmath}
\usepackage{amssymb}
\usepackage[T1]{fontenc}
\usepackage{etoolbox}
\usepackage{graphicx}
\usepackage{xcolor}
\usepackage{bigints}
\usepackage[section]{placeins}
\usepackage{float}
\usepackage{relsize}
\usepackage{color}
\usepackage{dcolumn}
\usepackage{bm}

\usepackage[height=10in, a4paper, hmargin={1.4cm,0.45in}]{geometry}

\begin{document}

 \title{Extended GUP formulation and the role of momentum cut-off}

 \author{Sebastiano Segreto}
 \email{sebastiano.segreto@uniroma1.it}
 \affiliation {Physics Department, 'Sapienza' University of Rome, P.le Aldo Moro 5, Rome, 00185, Italy}
 \author{Giovanni Montani}
 \email{giovanni.montani@enea.it}
  \affiliation {Physics Department, 'Sapienza' University of Rome, P.le Aldo Moro 5, Rome, 00185, Italy}
 \affiliation{ENEA, FSN-FUSPHY-TSM R.C. Frascati, Via E. Fermi 45, Frascati, 00044, Italy}

 \begin{abstract}
  \noindent We analyze the extension of the GUP theory deriving from the modified uncertainty principle in agreement with the string low energy limit, which represents one of the most general formulations satisfying the Jacobi identity, in the context of the associative algebras.
   After providing some physical insights on the nature of the considered approaches exploiting the cosmological arena, first, we show how a natural formulation of the theory in an infinite momentum space does not lead to the emergence of a nonzero minimal uncertainty in position, then we construct a truncated formulation of the theory in momentum space, proving that only in this case we can recover the desired feature of the presence of a nonzero minimal uncertainty in position, which - as usual in these theories - can be interpreted as a phenomenological manifestation of cut-off physics effects.
  Both quantization schemes are completely characterized and finally applied to study wave packets' behavior and their evolution in time. The obtained results can shed light on which generalizations of the GUP theory are more coherent with the string low energy limit, in view of the existence of a minimum length in the form of a minimal uncertainty in position.
 \end{abstract}	
                    
 \maketitle

 	\section{Introduction}
	
	Theories describing cut-off physics effects on the ultraviolet behavior of gravity, such as Loop Quantum Gravity 
	\cite{CQG:2014},\cite{Rovelli:2004tv} and Superstrings \cite{Polchinski:2014mva}, can have phenomenological representations equivalent to modified formulations 
	of non-relativistic quantum theory.
	If, on the one hand, the motion of particles in the low energy limit of string theories \cite{Amati:1987wq}-\nocite{Amati:1988tn}\nocite{Gross:1987ar}\nocite{Gross:1987kza}\cite{Konishi:1989wk} is well-described by the so-called Generalized Uncertainty Principle (GUP) theories,
	a quantum framework based on modified uncertainty relations between position and momentum operators, arising from a deformation of the Heisenberg algebra \cite{Kempf:1993bq}-\nocite{Kempf:1994qp}\cite{Kempf:1994su}, on the other, the proper application of Loop Quantum Gravity in the minisuperspace \cite{Ashtekar:2011ni}-\nocite{Corichi:2007tf}\cite{Barca:2021qdn} is perfectly summarized by 
	the quantum formalism known as Polymer Quantum Mechanics (PQM), which is essentially an implementation of the quantum theory on a lattice.
	
	   \noindent It is worth noting that, although these theories are in principle built on two different quantization schemes, recent works \cite{Barca:2021epy} suggest that the PQM formalism can be interpreted as a GUP theory as well, i.e. it is possible to faithfully represent it as a proper deformed algebra of quantum operators
	
	In this paper, we concentrate our attention on the first mentioned approach to cut-off physics, namely GUP theories.
    There exist a huge variety of generalizations of the original framework  discussed by Kempf, Mangano and Mann (KMM) in \cite{Kempf:1994su} 
    which have been widely studied in their structural aspects \cite{Bosso:2020aqm}-\nocite{Bosso:2021koi}\nocite{Bosso:2022rue}\cite{Bosso:2022vlz} and applied in several fields, from cosmology \cite{Battisti:2007jd}-\nocite{Battisti:2007zg}\nocite{Battisti:2008rv}\nocite{Battisti:2008qi}\nocite{Battisti:2008am}\cite{Battisti:2009at} 
    to black holes \cite{Nowakowski:2009ha}-\nocite{Arraut:2008hc}\cite{Arraut:2012yd}.
    Here, we completely re-analyze and precise some of the results discussed in \cite{Fadel:2021hnx} regarding one of these above mentioned generalizations, outlining some critical questions concerning the truncation of the	momentum space of the proposed theory.
    
	As it is known, one of the main issues and at the same time one of the most interesting features of these theories is the possible existence of a nonzero minimal uncertainty in position, 
	which can be naturally interpreted as a minimum length in the theory itself.
	Coherently with the basic idea of a string configuration, this is exactly the case for the GUP formulation studied in \cite{Kempf:1994su}, in which indeed the modified uncertainty principle
	comes directly from low energy considerations concerning the string theory (see above).
	In \cite{Maggiore:1993rv}-\nocite{Maggiore:1993kv}\cite{Maggiore:1993zu}, the possibility to generalize and extend the uncertainty principle mentioned above has been inferred by some heuristic observations, 
	leading to the introduction of a square root term in the Heisenberg algebra, able to reproduce the original approach in \cite{Kempf:1994su} in a proper limit. 
	
	This generalization acquires particular relevance when applied to the minisuperspace variables as discussed in \cite{Barca:2021epy} and \cite{Battisti:2008am}.
    In these works it is indeed shown how the semi-classical 
	model of the GUP theory at issue overlaps the (modified) Friedmann equation for an isotropic Universe typical of some models of \textit{brane cosmology} \cite{Randall:1999vf}.
    This relation can be considered as a potential mark for the physical interest of the theory itself.
    
	In the same works, it has been also emphasized how a simple change of sign in the square root allows switching to a polymer-like formulation, 
	associated with the (modified) Friedmann equation of Loop Quantum Cosmology \cite{Ashtekar:2015iza}, where the sign is semi-classically translated into the non-einsteinian correction of the dynamics. 
	
	As told before, many other generalizations of the analysis carried out in \cite{Kempf:1994su} have been considered over the years (see for example \cite{Gomes:2022} for an overview), but, as discussed in \cite{Barca:2021epy} and in \cite{Maggiore:1993rv}-\nocite{Maggiore:1993kv}\cite{Maggiore:1993zu}, the square root-modified Heisenberg algebra is the most general one - modulo a minus sign in the square root itself - that preserves the Jacobi identity of the operators and therefore the 
	only one providing a complete treatment, at least in the context of the associative algebras, the modification of which is obtained through the introduction of a function $f(\mathbf{\hat{p}})$, depending only on the momentum operator.
	
     It is worth mentioning that the non-associative algebras can represent an extremely interesting framework in which to implement the quantum theory (see for example \cite{liebmann2019non} and \cite{Dzhunushaliev:2005yd}), but we are not going to deal with them in this article.
	
	The analysis developed in \cite{Fadel:2021hnx} states, as one of the main conclusions, the existence of a minimum length in the form of a nonzero minimal uncertainty in position,
	as in the original approach in \cite{Kempf:1994su}. 
	In deriving this result a pivotal role is played by a series expansion of the square root term itself.
	
	It is exactly on this point that our analysis is focused, aiming to precise and clarify the conditions under which the Taylor expansion is allowed and to determine the proper implications
	of such a procedure.
	Indeed, in the case under study, the series expansion is mathematically viable only if the momentum space is restricted to a compact region, 
	needed to ensure the convergence of the series itself. 
	It is therefore clear that the conclusions exposed in \cite{Fadel:2021hnx} are not necessarily valid nor true in a complete, non-truncated formulation of the theory, which thus asks to be studied via different methods.
	On this ground, through a careful functional analysis of the position operator and by means of the techniques first developed and discussed in \cite{Detournay:2002fq}, in particular from considerations 
	on the divergence of the modified Lebesgue measure in the resulting
	Hilbert space, we preliminarily arrive to show that, actually, a zero minimal uncertainty in the position operator is predicted in the theory when the momentum space is not truncated.
	From this result we proceed to quantize this generalized scheme, outlining its intrinsic difference from the analysis carried out in \cite{Kempf:1994su}.
	Then, by exploiting again the method exposed in \cite{Detournay:2002fq}, we rigorously construct the quantum theory associated with 
	a truncated momentum space, characterizing completely the involved operators from a functional point of view, and showing how in this case and only in this case
	the existence of a nonzero minimal uncertainty in the position operator emerges, although slightly different from that one proposed in \cite{Fadel:2021hnx}. 
	This automatically allows us to construct a collection of maximally localized functions and therefore a quasi-position representation similar to that one first discussed in \cite{Kempf:1994su}.

	This means that the only viable generalization of the original GUP formulation in \cite{Kempf:1994su}, able to preserve the physical relevant fact of the existence of a minimum length, is the square root-modified one, but implemented through an \emph{ad hoc} truncation of the momentum space.
    This non-trivial aspect, namely the truncation, certainly calls attention to a possible physical justification.
   
    In that respect, cosmology, in particular in the minisuperspace formulation, offers an ideal arena to explore such a motivation, as we will precise in Section \ref{section_IV}.
    Indeed, we are going to show, by considering an isotropic Universe and by applying the GUP formalism to its volume-like variable, how it is possible to achieve in some measure, already at the classical level, a significantly different behavior of the dynamics with respect to the existence of the primordial singularity.

	The paper is structured as follows: in Section \ref{section_II}  we
	summarize and comment in some detail the functional procedure introduced first in \cite{Detournay:2002fq}, which will be fully exploited to construct the physical domain of the theory in the considered Hilbert space
	and to determine the maximally localized functions and their minimal uncertainty in position; in Section \ref{section_III} we introduce the extended GUP formulation obtained from the square root-modified Heisenberg algebra and we outline the analysis and the conclusion discussed in \cite{Fadel:2021hnx}; 
	in Section \ref{section_IV} we give some physical motivations to the structure of the formalism we are going to develop, based on some heuristic considerations in the cosmology arena, specifically regarding a minisuperspace model;
	in Sections \ref{section_V} and \ref{section_VI} we carry out our complete analysis 
	of this extended GUP formulation. In particular, in Section \ref{section_V} we construct and study the full theory, that is the theory implemented in a non-truncated momentum space, while in Section \ref{section_VI}, through the same steps, we construct and study the truncated or compact theory, that is the theory implemented 
	in a truncated momentum space.
	In the subsections \ref{section_VIa} and \ref{section_VIb} first we make a comparison of our results with the ones exposed in \cite{Fadel:2021hnx}, then, following the arguments presented in \cite{Kempf:1994su}
	on the possibility to recover information on the position, we construct the so-called quasi-position representation within our truncated theory.
	Finally, in Section \ref{section_VII}, we analyze the behavior of localized wave packets in both the truncated and non-truncated formulations, comparing the spreading properties with the ones resulting from
	the standard non-relativistic quantum mechanical approach and pointing out the relevant differences.
	In Sections \ref{section_VIII} we give our conclusions and summarize our results.

	\section{General GUP framework}
	\label{section_II}
	
    One of the possible rigorous ways to formulate a quantum theory with a modified uncertainty principle involving the position and momentum operators is the introduction of a modification in the Heisenberg algebra.
  Although this alteration can be realized in several ways, in the present paper we are going to take under exam algebras the structure of which can be realized as follows:
 \begin{equation} \label{general_algebra}
     \comm{\mathbf{\hat{x}}}{\mathbf{\hat{p}}}=i \hbar f(\mathbf{\hat{p}}),
 \end{equation}

 where for the function $f(p)$ we assume that:
 \begin{equation} \label{condition_zero}
     \exists \; C>0 \mid \forall p \in \mathbb{R} \quad  f(p)> C.
 \end{equation}

 The natural choice is to represent these algebras on momentum space, where the action of the operators, by means of the braket formalism, can be written as:
    \begin{align}
    	&\mathbf{\hat{x}}\ket{\psi} \rightarrow i \hbar f(p) \partial_{p} \psi(p) \\
    	&\mathbf{\hat{p}} \ket{\psi} \rightarrow p \psi(p), \quad p \in \mathbb{R}.
    \end{align}

	It is straightforward to verify that this representation satisfies the commutation relation.
	Nevertheless, it has to be clear that this is not the only possible choice.

    In order to construct our theory in a consistent manner, it is first strictly necessary to understand what conditions are imposed by the algebra on the operators and how these have to be properly defined.
    
	Restricting for simplicity to the one-dimensional case, as a first request, the $\mathbf{\hat{x}}$ and $\mathbf{\hat{p}}$ operators have to be defined in a  dense subspace of the Hilbert space of the theory, i.e. they have to be densely-defined, and in this domain they need to result to be closed and symmetric.
    To fulfill these demands, the Hilbert space in which we have to operate will be $\mathcal{L}^2\left(\mathbb{R}, \frac{dp}{f(p)}\right)$, where the modified Lebesgue measure must be introduced in order to make $\mathbf{\hat{x}}$ symmetric on its domain.
	Acting as a multiplicative operator, the momentum operator would be symmetric in any case, with respect to any measure.
	
	It is now crucial to understand if these operators are essentially self-adjoint operators, as the ordinary quantum theory asks.
	
	Indeed, as pointed out by Kempf et al. \cite{Kempf:1994su}, this mathematical property is the key to understanding what is really physically relevant in our analysis, that is to figure out whether there exists a minimum value in the position uncertainty $\Delta \mathbf{\hat{x}}$ different from zero and in that case which physical states realize it.

  For this purpose, we will now review and comment on a proposal for a more general approach, which can be found in \cite{Detournay:2002fq} (DGS method).
  According to this prescription, two cases need to be distinguished: the compact and the non-compact case.
  
  \subsection{The compact case}\label{subsection_I}
  
  Let us consider a function $f(\abs{p})$ such that, for $p\gg 1$, $f(\abs{p})\approx \abs{p}^{1+\nu}$, with $\nu>0$. In this case for the quantity:
  \begin{equation}
  	z(p)=\int_{0}^{p} f(q)^{-1} dq
  \end{equation}
  it is true that:
  \begin{equation} \label{conditions}
  	z(+\infty)= \alpha_{+} \; , \qquad z(-\infty)=\alpha_{-} \; , \quad \alpha_{\pm} \in \mathbb{R}.
  \end{equation}

 It is then possible to construct a diffeomorphism between $\mathbb{R}$ and the compact real interval $[\alpha_{-}, \alpha_{+}]$ through the map $p \to z(p)$, moving on from the space $\mathcal{L}^2(\mathbb{R}, dp/f(p))$ to the space $\mathcal{L}^2([\alpha_{-}, \alpha_{+}], dz )$.
 
 On our new Hilbert space, the $\mathbf{\hat{z}}$ operator is a symmetric multiplicative operator defined on the whole Hilbert space, hence, it is automatically self-adjoint and, due to the Hellinger-Toeplitz theorem, it is bounded.
 
 \noindent The same can be stated for  $\mathbf{\hat{z}^2}$.
 
 \noindent With regards to the $\mathbf{\hat{x}}$ operator, the most natural choice is to define it as follows:
  \begin{align}
 	\mathbf{\hat{x}} \; : \quad \! \!  \mathcal{D}_{\mathbf{\hat{x}}} \; &\longrightarrow \; \mathcal{L}^2 (  [\alpha_{-}, \alpha_{+}], dz ) \\
 	 \psi & \; \mapsto \; i \hbar \partial_z^{(w)} \psi,
  \end{align}
 where \footnote{By $\mathcal{H}^{1,2}$ we denote the space known as first Sobolev space, that is the space of all square-integrable functions, which first weak derivatives are also square-integrable.
 The norm of the space is defined as:$\norm{f}^2=\int_{\alpha_-}^{\alpha_+} (\abs{f(z)}^2+\abs{f'(z)}^2) dz$}: 
  \begin{equation} 
  	\begin{split}
  	\mathcal{D}_{\mathbf{\hat{x}}}=&\{\psi(z) \in \mathcal{H}^{1,2}([\alpha_{-}, \alpha_{+}], dz) \mid \\
  	& \quad \! \psi(\alpha_{-})=\psi(\alpha_{+})=0 \}.
  	\end{split}
  \end{equation}
 and the symbol $\partial^{(w)}$ stands for the weak derivative or the derivative in the distributional sense.
 
 This operator is not self-adjoint as it can be seen through a direct construction of its adjoint, which turns out to be a true extension, that is $\mathbf{\hat{x}} \subsetneq \mathbf{\mathbf{\hat{x}}^{\dagger}}$.
 
 In particular:
 \begin{align} 
 	\mathbf{\hat{x}}^{\dagger} \;: \quad \!\! \mathcal{D}_{\mathbf{\hat{x}}^{\dagger}} \; & \longrightarrow  \; \mathcal{L}^2([\alpha_{-}, \alpha_{+}], dz)\\
 	\psi&  \; \mapsto \; i \hbar \partial_{z}^{(w)} \psi,
 \end{align}
 where:
 \begin{equation}
 	\mathcal{D}_{\mathbf{\hat{x}}^{\dagger}} = \{\psi(z) \in \mathcal{H}^{1,2}([\alpha_{-}, \alpha_{+}], dz) \}.
 \end{equation}
 
 Being the adjoint operator well-defined, at this point it is possible to calculate the deficiency indices $(d_{+}, d_{-})$ of $\mathbf{\hat{x}}^{\dagger}$, i.e. the dimension of the kernel of the operators $(\mathbf{\hat{x}}^{\dagger} \pm i \mathbb{\mathbf{I}})$.
 In the z-representation, this will lead to the following differential equations:
 \begin{equation} \label{def_indices_zeta}
    	\begin{split}
    	(\mathbf{\hat{x}}^{\dagger} \pm i \mathbb{\mathbf{I}})\ket{\psi}&=0 ,\\
    	    \partial_{z}^{(w)}\psi(z) \pm  \psi(z) &= 0, \\
    	\partial_{z}^{(w)} \left(e^{\pm \frac{1}{\hbar} z} \psi(z) \right)&=0.
    	\end{split}
    \end{equation}
   Since the function on which the weak derivative acts is a locally integrable function, it is possible to conclude, 
    by an application of the De Bois - Reymond lemma, 
   that it is equal to a constant $\kappa$ almost everywhere on $[\alpha_{-}, \alpha_{+}]$, hence:
    \begin{equation}
    	\psi(z)=\kappa  e^{\mp \frac{z}{\hbar}}.
    \end{equation}
  As it is straightforward to verify, both functions belong to $\mathcal{D}_{\mathbf{\hat{x}}^{\dagger}}$, therefore in this case the deficiency indices will be $(d_{+}, d_{-})=(1,1)$. According to the Von Neumann's theorem, this means that  $\mathbf{\hat{x}}$ is not essentially self-adjoint, but rather it admits a one-parameter family of self-adjoint extensions.

  We notice that in p-representation, the previous differential equations to be solved would be:
  \begin{equation} \label{def_indices}
    	\partial_{p}^{(w)} \left(e^{\pm \frac{1}{\hbar} \int { dp f(p)^{-1}}} \psi(p) \right)=0,
    \end{equation}
   the solutions of which, by the same arguments, are:
    \begin{equation}
    	\psi(p)=\kappa  e^{\pm \frac{1}{\hbar} \int { dp f(p)^{-1}}}.
    \end{equation}

 From these expressions the pivotal role played by the function $f(p)$ and therefore, in some respect, by the algebra itself, appears explicitly.
 Indeed, as we have just seen, it is the fact that $f(p)$ satisfies the conditions \eqref{conditions} that has led to obtain a position operator which allows infinitely many self-adjoint extensions.
    
 These self-adjoint extensions $\mathbf{\mathbf{\hat{x}}_{\lambda}}$ have the same action of  $\mathbf{\hat{x}}$ but are defined on the domain: 
  \begin{equation}
  	\begin{split}
  		\mathcal{D}_{\mathbf{\mathbf{\hat{x}}_{\lambda}}}=\{&\psi(z) \in \mathcal{H}^{1,2}([\alpha_{-}, \alpha_{+}], dz) \mid \\  &  \psi(\alpha_{+})=e^{-i\lambda}\psi(\alpha_{-})\}, \qquad \lambda \in \mathbb{R}.
  	\end{split}
  \end{equation}

 Not surprisingly, neither the squared position operator $\mathbf{\mathbf{\hat{x}}^2}$ is essentially self-adjoint, but admits a one-parameter family of self-adjoint extensions given by $\mathbf{\mathbf{\hat{x}}^2_{\lambda}}$.
 Nevertheless, as it is extensively discussed in \cite{Detournay:2002fq}, the more convenient choice for the construction of the squared position operator is the operator $\mathbf{\mathbf{\hat{x}}^{\dagger}\mathbf{\hat{x}}}$, which domain is exactly	$\mathcal{D}_{\mathbf{\hat{x}}}$.
 Clearly, since $\mathbf{\mathbf{\hat{x}}^2} \subsetneq\mathbf{\hat{x}}$, $\mathbf{\mathbf{\hat{x}}^{\dagger}\mathbf{\hat{x}}}$ is an extension of $\mathbf{\mathbf{\hat{x}}^{2}}$ and it results to be self-adjoint.
 Our final list of operators will be then represented by $\mathbf{\hat{x}}_{\lambda}, \mathbf{\mathbf{\hat{x}}^{\dagger}\mathbf{\hat{x}}},\mathbf{\hat{z}}, \mathbf{\hat{z}}^2$.
 
 We can now define the domain of the commutator $\comm{\mathbf{\hat{x}}_\lambda} {\mathbf{\hat{z}}}$, which turns out to be:
  \begin{equation}
 		\mathcal{D}_{\comm{\mathbf{\hat{x}}_\lambda}{\mathbf{\hat{z}}}}= 	\mathcal{D}_{\mathbf{\hat{x}}_{\lambda}\mathbf{\hat{z}}} \cap \mathcal{D}_{\mathbf{\hat{z}}\mathbf{\hat{x}}_{\lambda}}=
 		\mathcal{D}_{\mathbf{\hat{x}}_{\lambda}\mathbf{\hat{z}}} \cap \mathcal{D}_{\mathbf{\hat{x}}_{\lambda}}= \mathcal{D}_{\mathbf{\hat{x}}}.
  \end{equation}
 
 Finally we are able to define the \textit{physical space} of the theory.
 In the general case, as physical space we can consider that subspace of the Hilbert space resulting from the intersection between the domains of the operators $\mathbf{\hat{x}}, \mathbf{\hat{p}}, \mathbf{\hat{x}}^2, \mathbf{\hat{p}}^2$ and $[\mathbf{\hat{x}},\mathbf{\hat{p}}]$, in order to make it possible to define the uncertainty in position and momentum and so that the generalized uncertainty principle (GUP) holds.
 
 Therefore, in the studied case, this space, which is the space where to look for the maximally localized states as well, will be:
  \begin{equation}
 		\mathcal{D}_{\mathbf{\hat{x}}_{\lambda}} \cap \mathcal{D}_{\mathbf{\mathbf{\hat{x}}^{\dagger}\mathbf{\hat{x}}}} \cap \mathcal{D}_{\mathbf{\hat{z}}} \cap \mathcal{D}_{\mathbf{\hat{z}^2}} \cap \mathcal{D}_{\comm{\mathbf{\hat{x}}_\lambda}{\mathbf{\hat{z}}}}=\mathcal{D}_{\mathbf{\hat{x}}}.
 \end{equation}
 
 At this point it is possible to write down $\Delta {\mathbf{\hat{x}}}$ as a functional object and define the maximally localized states as those states which minimize this expression in $\mathcal{D}_{\mathbf{\hat{x}}}$:
 	\begin{align} \label{functional_uncertainty}
 		(\Delta {\mathbf{\hat{x}}})^2&=\frac{\bra{\Psi}(\mathbf{\mathbf{\hat{x}}^{\dagger}\mathbf{\hat{x}}}-\xi)\ket{\Psi}}{\bra{\Psi}\ket{\Psi}}, \quad \ket{\Psi} \in \mathcal{D}_{\mathbf{\hat{x}}},\\
 		(\Delta {\mathbf{\hat{x}}}^{min})^{2}&=\text{min}\frac{\bra{\Psi}(\mathbf{\mathbf{\hat{x}}^{\dagger}\mathbf{\hat{x}}}-\xi)\ket{\Psi}}{\bra{\Psi}\ket{\Psi}} := \mu ^2.
 	\end{align}
 
 Two constraints must be imposed for a consistent procedure:
 \begin{align}
 	&\xi=\frac{\bra{\Psi}\mathbf{\hat{x}}_{\lambda}\ket{\Psi}}{\braket{\Psi}}, \quad \xi \in \mathbb{R},\\
 	&\gamma=\frac{\bra{\Psi}v(\mathbf{\hat{p}})\ket{\Psi}}{\braket{\Psi}}, \quad  \gamma \in \mathbb{R}.
 \end{align}
 
 The first one concerns the existence of the expectation value of $\mathbf{\hat{x}}_{\lambda}$, while the second one, which plays a crucial role in the whole method, is an extra condition in order to select only those states which admit a finite expectation value of some function of the momentum operator, e.g. the energy, from a physical point of view.
 The constrained variational principle gives back, in p-representation, the Euler-Lagrange equations for the system:
 \begin{equation} \label{EL_eq}
 	\begin{split}
    \biggl\{&-\left[f(p) \partial_{p}^{(w)}\right]^{2}-\xi^{2}+2 a\left(i f(p) \partial_{p}^{(w)}-\xi\right) \\
    &+ 2 b(v(p)-\gamma)-\mu^
{2}\biggr\} \Psi(p)=0.
    \end{split}
 \end{equation}
 Since $v(p)$ is an arbitrary function, it is impossible to write down a general solution of \eqref{EL_eq}, except for the case $b=0$.
 For this particular choice of the Lagrange multiplier, taking into account the first constraint and the boundary conditions of the domain, the solution eventually will be:
 \begin{align}
 	&\Psi(p)=\mathcal{C} \exp [-i \xi z(p)] \sin{\{\mu\left[z(p)-\alpha_{-}\right]\}}, \label{general_max_Loc_states} \\
 	&\mathcal{C}=\sqrt{\frac{2}{\alpha_{+}-\alpha_{-}}} \;, \quad \mu=\frac{ n \pi}{\alpha_{+}-\alpha_{-}} \, , \, n \in \mathbb{N}_{0}.
  \end{align}

 This function correctly belongs to $\mathcal{D}_{\mathbf{\hat{x}}}$ and it has a non-vanishing minimal uncertainty in position, obtained for $n=1$, which corresponds to the minimum of the functional \eqref{functional_uncertainty}, for $b=0$.
 
 Of course, the value $b=0$ is a special case. Yet, it can be shown that $({\Delta{\mathbf{\hat{x}}}}^{min})^{2}\mid_{b=0}$ is a local minimum with respect to $\gamma$, hence signaling the importance of such a solution (for a detailed discussion see \cite{Detournay:2002fq}). 
 
  We stress that this method for searching an absolute minimum value of $\Delta \mathbf{\hat{x}}$ is more general with respect to 
  the one discussed in \cite{Kempf:1994su}.
  Indeed, first of all, the latter is valid only for some class of functions $f(p)$ associated with the operator $f(\mathbf{\hat{p}})$ of the deformed algebra and secondly, its more evident limit is to look for maximally localized states only among the so-called squeezed states, that is those states for which the Heisenberg inequality is saturated:
  \begin{equation} \label{HUP}
    	\Delta{\mathbf{\hat{x}}} \Delta{\mathbf{\hat{p}}} = \frac{1}{2} \abs{\expval{\comm{\mathbf{\hat{x}}}{\mathbf{\hat{p}}}}}.
    \end{equation}

   Even if successful in the case studied by Kempf et al., for a general algebra such as $\comm{\mathbf{\hat{x}}}{\mathbf{\hat{p}}}=i \hbar f(\mathbf{\hat{p}})$ it is not obvious at all the exhaustiveness of such a research. The most intuitive way to understand it is to notice that the expectation value of the commutator, in the general case, is state-dependent.
  Thus it is clear that it is necessary to look for these maximally localized states in a more wide domain of our Hilbert space, which is exactly the point addressed by the method that we have summarized above.
  This means that, although it is still possible that the states \eqref{general_max_Loc_states} are squeezed states, this will not be the general case anymore.

 \subsection{The non-compact case}
 \label{subsection_II}
  
  If the quantity $z(p)$ diverges for $p \to \infty $, the previous procedure, which is essentially based on the mapping between $\mathbb{R}$ and a real compact interval, is not available. In general this happens when $f(p)\approx \abs{p}^{1+ \nu } $, for $ \abs{p} \to \infty$, with $ \nu <0 $.
  This suggests that for these GUP theories there is no an absolute minimal uncertainty in position different from zero.
  In order to show that this is indeed the case, we need to turn our attention to the squeezed states of these theories. 
  Indeed in these scenarios the squeezed states are physical states, in the sense fixed by the constrained variational principle of the previous section, but their uncertainty in position can be made arbitrarily small.
  This leads to conclude that the absolute minimum of the quantity $\Delta{\mathbf{\hat{x}}}$ is (asymptotically) zero in these theories. 
  
  To prove that this is exactly the case and since it will be useful for the next sections, now we are going to briefly summarize the procedure through which it is possible to determine these squeezed states and their uncertainty in position, following once again what is discussed in \cite{Detournay:2002fq}.
  
  As it is known, relation \eqref{HUP} can be obtained directly from the algebra, under the minimal assumptions that the operator $\mathbf{\hat{x}}$ and $\mathbf{\hat{p}}$ are dense and symmetric operators on their domain of definition.
	Tracing back all the steps from \eqref{general_algebra} to \eqref{HUP}, it can be shown that the equality sign in \eqref{HUP} can be obtained from those states that are eigenstates of null eigenvalue of the following operator:
	\begin{equation}
		\mathcal{\hat{A}}_{\Lambda}\ket{\Psi_\Lambda} := (\mathbf{\hat{x}}-\xi)+ i \hbar \Lambda (\mathbf{\hat{p}}-\eta)\ket{\Psi_\Lambda}=0,
	\end{equation}
	where $\xi,\eta,\Lambda$ are real parameters.
	
	In p-representation, this is a first-order differential equation the solution of which is:
	\begin{equation}
		\Psi_\Lambda(p) =\mathcal{N}e^{\frac{(\hbar \Lambda \eta - i \xi)}{\hbar}z(p)-\Lambda u(p)},
	\end{equation}
	where 
	\begin{align} 
	&z(p) = \int_{0}^{p}  dq f(q)^{-1}\label{z(p)}, \\
	&u(p)=\int_{0}^{p}  dq \;  q f(q)^{-1} \label{u(p)},
    \end{align} 
    and $\mathcal{N}$ is just a normalization constant.
    
	Clearly, the obtained wave function must be a square-integrable function, or, in other words, it has to be normalizable.
	Once $\Lambda$ is fixed to be strictly positive, this condition is satisfied  whenever:
	\begin{equation}
	 \lim_{p \to \pm \infty} u(p)=+\infty.
	\end{equation}
	
	Furthermore, the fulfillment of the above requirement assures that the parameters $\eta$ and $\xi$ coincide with the expectation values of $\mathbf{\hat{x}}$ and $\mathbf{\hat{p}}$, as they should.
	It is relevant to notice that this last condition is related to the need of working with a symmetric position operator. 
	
	Now, in order to obtain the relevant quantities we are interested in, namely $\Delta \mathbf{\hat{x}}_{\Psi_{\Lambda}}$ and $\Delta \mathbf{\hat{p}}_{\Psi_{\Lambda}}$, we need to evaluate and properly compare two different objects.
	On the one hand, the norm of the state $\mathcal{\hat{A}}_{l}\ket{\Psi_{\Lambda}}=\ket{\mathcal{\hat{A}}_{l} \Psi_\Lambda}$, where in general $l \neq \Lambda$:
	\begin{equation} \label{norm_sq}
		\begin{split}
		&\braket{\Psi_\Lambda \mathcal{\hat{A}}_{l} }{\mathcal{\hat{A}}_{l} \Psi_\Lambda}=\left(\bra{\Psi_\Lambda}\mathcal{\hat{A}}^{\dagger}_{l} \right)\left(\mathcal{\hat{A}}_{l}\ket{\Psi_\Lambda}\right)\\
		&= \hbar^2(l-\Lambda)^2\bigg( \bra{\Psi_\Lambda}(\mathbf{\hat{p}}^{\dagger}-\eta)\bigg)\bigg((\mathbf{\hat{p}}-\eta)\ket{\Psi_\Lambda}\bigg)\\
		&=\hbar^2(l-\Lambda)^2 \bra{\Psi_\Lambda}(\mathbf{\hat{p}}-\eta)^2\ket{\Psi_\Lambda},
		\end{split}
	\end{equation}
    where in the last line we have used the fact that $\big(\bra{\Psi_{\Lambda}}\mathbf{\hat{p}}^{\dagger}\big)\big(\mathbf{\hat{p}}\ket{\Psi_{\Lambda}}\big)=\bra{\Psi_{\Lambda}}\big(\mathbf{\hat{p}}^2\ket{\Psi_{\Lambda}}\big)$.
    On the other, we need to focus on the following quantity: 
	\begin{equation} \label{shift_norm_sq}
		\begin{split}
		&\bra{\Psi_\Lambda}\! \bigg(\mathcal{\hat{A}}^{\dagger}_{l}\mathcal{\hat{A}}_{l}\ket{\Psi_{\Lambda}}\bigg)\\
		&=\bra{\Psi_\Lambda}\! \bigg(\! (\mathbf{\hat{x}}-\xi)^2\! - \! \hbar l f(\mathbf{\hat{p}})\!+ \! \hbar^2 l^2(\mathbf{\hat{p}}\!-\! \eta)^2 \!  \ket{\Psi_{\Lambda}}\!\! \bigg),
		\end{split}
    \end{equation}
  the computation of which instead relies basically on the fact that $\bra{\Psi_{\Lambda}}\big(\mathbf{\hat{x}}^{\dagger}\mathbf{\hat{x}}\ket{\Psi_{\Lambda}}\big)=\bra{\Psi_{\Lambda}}\big(\mathbf{\hat{x}}^2\ket{\Psi_{\Lambda}}\big)$.
  
   Notice that we have stressed the use of parenthesis since it is relevant to understanding on which side the operators act.
   
   These two last expressions \eqref{norm_sq} and \eqref{shift_norm_sq} are equal if and only if:
   \begin{equation} \label{symm_cond_x^2}
   	\lim_{p \to \pm \infty} p\abs{\Psi_{\Lambda}}^2= 0,
   \end{equation}
   This condition, which will depend on the behavior of $u(p)$, comes essentially from the requirement that $\big(\bra{\Psi_{\Lambda}}\mathbf{\hat{x}}^{\dagger}\big)\big(\mathbf{\hat{x}}\ket{\Psi_{\Lambda}}\big)=\bra{\Psi_{\Lambda}}\big(\mathbf{\hat{x}}^2\ket{\Psi_{\Lambda}}\big)$.
  
  Whenever these conditions are satisfied, we are allowed to compare \eqref{norm_sq} and \eqref{shift_norm_sq} and to write:
  \begin{align} 
  		(\Delta \hat{\mathbf{x}})_{\Psi_\Lambda}^{2}(\Delta \hat{\mathbf{p}})_{\Psi_\Lambda}^{2} &=\frac{\hbar^2}{4}\langle f(\hat{\mathbf{p}})\rangle_{\Psi_\Lambda}^{2}, \\
  		\Lambda &=\frac{\langle f(\hat{\mathbf{p}})\rangle_{\Psi_\Lambda}}{2(\Delta \hat{\mathbf{p}})_{\Psi_\Lambda}^{2}} \label{identif_DGS_Lambda},\\
  		(\Delta \hat{\mathbf{x}})_{\Psi_\Lambda}^{2} &=\frac{\hbar^2}{4} \frac{\langle f(\hat{\mathbf{p}})\rangle_{\Psi_\Lambda}^{2}}{(\Delta \hat{\mathbf{p}})_{\Psi_\Lambda}^{2}}=\hbar^2\Lambda^{2}(\Delta \hat{\mathbf{p}})_{\Psi_\Lambda}^{2} \label{Min_Deltax_DGS},
  \end{align}

  whit a clear meaning of the used symbols.

  The equation \eqref{Min_Deltax_DGS} represents an explicit functional expression for $\Delta \mathbf{\hat{x}}_{\Psi_\Lambda}$:
  \begin{equation} \label{functional_Deltax}
 	(\Delta \hat{\mathbf{x}})_{\Psi_\Lambda}^{2}\!\! \! \!=\!\hbar^2 \Lambda^{2} \frac{\bigints_{-\infty}^{+\infty}\frac{\left(\! p^{2}\!-\!\eta^{2}\!\right) exp[-2 \Lambda(u(p)-\eta z(p))]} {f(p)} d p}{\bigints_{-\infty}^{+\infty} \frac{exp[-2 \Lambda(u(p)-\eta z(p))]} {f(p)} d p}.
  \end{equation}
  By minimizing this object with respect to $\eta$ and $\Lambda$, it is in principle possible to determine the value of $\Delta \mathbf{\hat{x}}^{min}$ and hence the wave function which realizes this uncertainty in position. 
  
  All the conditions discussed above are satisfied whenever the function $f(p)$ grows at $\abs{p} \to \infty$ as $\abs{p}^{1+\nu}$, with $-1 \leq \nu \leq 1$, where in particular the lower bound is needed to assure the validity of the condition \eqref{condition_zero}.
  Therefore for the case that we are considering, that is $\nu<0$, the procedure holds and it can be shown that, for $\Lambda \to 0$:
  \begin{equation}
      (\Delta \hat{\mathbf{x}})_{\Psi_\Lambda}^{2} \propto \Lambda^{\frac{2 \nu}{\nu-1}} \to 0
  \end{equation}
  which proves that for these GUP theories there is no minimal length in the guise of a non-zero minimal uncertainty in position.

 \section{Extended GUP formulation} 
  \label{section_III}
  The literature is plenty of GUP-modified frameworks built as an extension or a generalization of the KMM one, which reads as:
  \begin{equation}
      \comm{\mathbf{\hat{x}}}{\mathbf{\hat{p}}}=i \hbar( 1+\beta \mathbf{\hat{p}}^2),
  \end{equation}
  where $\beta$ is a dimensional deformation parameter, which, to some extent, controls the deviation from the standard commutator\footnote{In these theories the $\beta$ parameter, from which the value of the possible minimum length will depend, is always treated as a constant. Nevertheless there exist some formulations of string theory, namely dynamical string tension theories, the low energy limit of which could result in something different, such as perhaps a variable $\beta$ and hence a not universal minimum length (see e.g. \cite{Guendelman:2021zmf}).}.

  It can be noticed at first glance that the KMM-modified commutator can be regarded as a perturbative expansion in $\beta$ (at the first order) of some more general operator-valued function.
  
  One of the most interesting cases is doubtless represented by the GUP theory studied in \cite{Fadel:2021hnx}, \cite{Maggiore:1993kv}, \cite{Maggiore:1993zu}, the deformed algebra of which can be written as:
  \begin{equation} \label{MM_GUP}
  	\comm{\mathbf{\hat{x}}}{\mathbf{\hat{p}}}=i \hbar \sqrt{1+2 \beta \mathbf{\hat{p}}^2}.
  \end{equation}
  
 This specific modification of the canonical commutation relation (CCR) has a high degree of generality.
 The reason for that lies in the fact that, in a 3-dimensional setting, the most general modified algebra that can be written, asking that the groups of translations and the group of rotations be undeformed, is:
\begin{align} 
 	\comm{\mathbf{\hat{x}}_{i}}{\mathbf{\hat{x}}_{j}}=&\frac{\hbar}{\kappa c^2} a(\mathbf{\hat{p}}) \epsilon_{ijk} \mathbf{\hat{J}}_{k},  \label{generalized_commm1} \\
 	\comm{\mathbf{\hat{x}}_{i}}{\mathbf{\hat{p}}_{j}}=& \, i\hbar \delta_{ij} f(\mathbf{\hat{p}}).  \label{generalized_commm2} 
 \end{align}
  where $a(p)$ and $f(p)$ are sufficiently regular functions of the momentum operator,  $\mathbf{\hat{J}}$ is the angular momentum operator, $\kappa$ is a parameter with the dimension of a mass and $c$ is the speed of light in vacuum.
  By imposing, as it should be, that the constructed algebra satisfies the Jacobi identities, a differential equations system is obtained for the form of $a(p)$ and $f(p)$, the most general solution of which states that $a(p)=\pm1$, previous a rescaling of $\kappa$, and consequently $f(p)=\sqrt{\alpha \pm p^2/(\kappa c^2)}$, where $\alpha$ is an integration constant.
  By choosing $\alpha=1$ in order to recover in the proper limit the standard Heisenberg uncertainty principle and rewriting $1/(\kappa^2 c^2)=2\beta_0/(\mathcal{M}_{pl}^2 c^2)=2 \beta$ to make contact with the KMM notation - where $\beta_0$ is a dimensionless constant and $\mathcal{M}_{pl}$ is the Planck mass - in the end we can write $f(p)=\sqrt{1 \pm 2 \beta p^2}$.
  Therefore, the modified algebra \eqref{MM_GUP}, is one of the two general solutions obtained by imposing that the fundamental requirement of Jacobi identities is fulfilled by the general commutators \eqref{generalized_commm1}-\eqref{generalized_commm2}.
  
  An analysis carried out in \cite{Fadel:2021hnx} shows how this formulation of the quantum theory leads to the existence of a minimum length, namely a nonzero minimal uncertainty in position.
  By following the reasoning path of the authors and restricting for simplicity to the one-dimensional case, from \eqref{MM_GUP} the uncertainty principle can be derived:
  \begin{equation}
  	\Delta{\mathbf{\hat{x}}}\Delta{\mathbf{\hat{p}}} \geq \frac{\hbar}{2} \expval{\sqrt{1+2\beta\mathbf{\hat{p}}^2}}.
  \end{equation}  
  In order to evaluate explicitly the expectation value of the commutator, a series expansion is performed:
  \begin{equation} \label{series_exp_comm}
  	\sqrt{1+2\beta\mathbf{\hat{p}}^2}=\sum_{n} c_{n} (2\beta\mathbf{\hat{p}}^2 )^n,
  \end{equation}
  where $c_n$ is the generalized binomial coefficient:
  \begin{equation}
  	c_n=\binom{1/2}{n}=\frac{(-1)^2 (2n)!}{2^{2n}(1-2n)(n!)^2}.
  \end{equation}
  Then, after a brief chain of inequalities, it is possible to write:
  \begin{equation}
  	\Delta{\mathbf{\hat{x}}}\Delta{\mathbf{\hat{p}}} \geq \frac{\hbar}{2}\sqrt{1+ 2 \beta (\Delta{\mathbf{\hat{p}}})^2}.
  \end{equation}
   From this it is immediate to see that there exists indeed a nonzero minimum value for $\Delta\mathbf{\hat{x}}$, which is reached asymptotically, that is for $\Delta{\mathbf{\hat{p}}}$ approaching infinity.
   In particular:
   \begin{equation}
   	\Delta{\mathbf{\hat{x}}}^{min}=l_{P}\sqrt{\frac{\beta_0}{2}},
   \end{equation}
  where $l_{P}$ is the Planck length. 
  So according to \cite{Fadel:2021hnx}, this framework provides the theory with a "natural" minimum length and it is capable of doing so in a way that resembles much closer to the ordinary quantum theory with respect to the KMM GUP, due to the asymptotic behavior discussed above. 
  
  Nevertheless this procedure suffers from a subtle problem.
  The operator-valued function $\sqrt{1+2\beta\mathbf{\hat{p}}^2}$ admits a series expansion which is convergent if and only if the $\mathbf{\hat{p}}$ operator has a finite norm, in particular if and only if $\norm{\mathbf{\hat{p}}}\leq 1/\sqrt{2\beta}$.
  This means that the procedure discussed in \cite{Fadel:2021hnx} cannot be valid for the "usual" momentum operator, which is an unbounded operator, but it holds only for a theory where the momentum operator is properly bounded.
  If this is the case, it is clear that also $\Delta{\mathbf{\hat{p}}}$ will be a bounded quantity, therefore it does not make any sense to explore indefinitely the   $\Delta{\mathbf{\hat{p}}}$-region, since it will be accessible only up to a certain finite positive value. 
  From these considerations, it is clear that other paths are needed to properly explore this GUP theory and thus address these problems which make the conclusion unreliable.
  
  Two frameworks will be developed and studied: the full or non-truncated theory and the compact or truncated theory.
  
  A structure of this kind, even if on a formal level appears in a natural way during the analysis itself, as we will see, on a physical ground it stands in need of some motivations.
  
  The next section is dedicated to the attempt of addressing, in some measure, this point.

  \section{Heuristics and motivations}
  \label{section_IV}
As we discussed in some detail in the previous sections, the possibility to generalize the standard GUP approach described in \cite{Kempf:1994su} by introducing a square root term on the right-hand side of the commutator is mathematically well justified in view of the validity of the Jacobi identities.
The first intuitive physical interpretation is that of a much more general setting for the study of the properties of such theories, which reduces 
to the standard one if the momentum is much smaller than the value $\sim 1/\sqrt{\beta}$ (i.e. the square root is truncated up to 
the leading two orders of approximation). 

The matter concerning the two different physical situations, viz the truncation 
and non-truncation of the momentum space, is, on the other hand, a bit more delicate and subtle.
As we clearly stated in the introduction and as we are going to see in a rigorous manner, these two schemes of the extended Heisenberg algebra are 
not equivalent since only the truncated scenario is able to reproduce a minimal uncertainty in the position or in 
the generalized coordinate. This property can be considered the distinctive characteristic of a GUP approach, given its deep physical meaning and its consequences.   

In order to understand how the distinction between the two theories is relevant from a dynamical point of view and therefore the relative physical implications, we now turn our attention to a Minisuperspace model, namely the implementation of the generalized GUP framework 
to the dynamical variables associated with the description of a homogeneous and isotropic Universe. 
Indeed, it is just in the cosmology arena, when the generalized coordinates correspond to the Universe scale factors (in the example 
below we limit our attention to the volume of the isotropic Universe) that the physics provided by the deformation of the uncertainty principle becomes relevant,
since it is able to ensure significant insights about the primordial evolution of the Universe itself near the Planck era. 
In fact, we can consider the deformation parameter $\beta$ as a quantity in which the information about a quantum modification of the dynamics, which takes place only when the Universe density approaches the planckian value, is stored.

In what follows, for our heuristic purpose, we will mainly concentrate our attention on the classical dynamics of the system as induced by modified Poisson brackets, nevertheless the \emph{correspondence principle} immediately suggests that a restated classical dynamics (\emph{de facto} a modified gravity theory) is to be regarded as the phenomenological emergence of an intrinsically deformed quantum canonical dynamics of the primordial Universe \cite{CQG:2014}.

In this respect, let us consider the dynamics of a typical interesting paradigm for the primordial Universe, corresponding to the isotropic flat  Robertson-Walker model, in the presence of a free massless scalar field $\phi$. If we use as metric variable the 
cubed scale factor $v$, that is the variable that scales the comoving volumes, the action of this system reads as:

\begin{equation}
	S=\int_{t1}^{t2} dt \left\{ p_v\dot{v}+p_{\phi}\dot{\phi} - \frac{N c}{4 \pi^2} \left(-\frac{3\chi vp_v^2}{2} + \frac{p_{\phi}^2}{v}\right)\right\}\, , 
	\label{action_miniss}
\end{equation}
where $\chi$ is the Einstein constant, $c$ is the speed of light in the vacuum, $p_v$ and $p_{\phi}$ denote the conjugate momenta to the variables $v$ and $\phi$, respectively,  while $N$ is the lapse function and the dot refers to differentiation with respect to the coordinate time $t$.

As it is known, the variation with respect to $N$ will produce a dynamical constraint (reflecting the time covariance): 
\begin{equation}
	p_v^2 = \frac{2}{3\chi}\frac{p_{\phi}^2}{2v^2} \equiv \frac{2}{3\chi}\rho_{\phi}
	\, , 
	\label{constraint_miniss}
\end{equation}

where we denoted by $\rho_{\phi}$ the scalar field energy density. 

Since the scalar field is typically used in cosmology as a physical clock, we are naturally led to implementing the generalized GUP framework on the variables
$v$ and $p_v$ only, via the modified Poisson bracket:
\begin{equation}
	\left\{ v,\, p_v\right\} = \sqrt{1 + 2 \beta p_v^2}\, . 
	\label{PB_mod}
\end{equation}

As shown in \cite{Barca:2021epy} and \cite{Battisti:2008am}, the classical dynamics so recast is associated with a generalized Friedmann equation \cite{Montani:2009hju}, which corresponds exactly to that one dictated by some models of brane cosmology \cite{Randall:1999vf}. 

Let us now briefly derive this result.

The first modified Hamilton equation reads, in a synchronous reference frame ($N\equiv 1$), as follows:
\begin{equation}
	\dot{v} = - \frac{c}{\pi^2}\frac{3}{4}\chi v p_v \sqrt{1 + 2 \beta p_v^2}\, . 
	\label{v_eq_GUP}
\end{equation}

By observing that the Hubble rate $H$ takes the simple form $H = \dot{v}/3v$ in the volume variable and by making use of constraint \eqref{constraint_miniss}, 
it is straightforward to arrive at the desired modified Friedmann equation:
\begin{equation}
	H^2 =  \frac{c^2}{\pi^4} \frac{\chi}{24}\rho_{\phi}\left(	1 + \frac{\rho_{\phi}}{\rho^*}\right) \, , 
	\label{H_eq_GUP}
\end{equation}
where 
$\rho^{*}=3\chi/4\beta$ is a constant representing the characteristic energy density of the model.

The equation above is exactly the same one that can be obtained in the Randall-Sundrum model of brane cosmology, where 
$\rho^*$ depends on the brane tension $\sigma$ instead of $\beta$.
This equation contains a quadratic dependence on the energy density of the field and predicts the presence of a singularity where the Universe's volume 
vanishes.

On this behalf, we are naturally led to infer that our generalized GUP paradigm can have, in the Minisuperspace, a phenomenological relation with the brane physics. More precisely, while in the physical space the original GUP formulation reproduces features of the low energy string dynamics (see above for the references), 
in the configurational space of cosmology the same correspondence can be drawn between the generalized square root GUP physics and the braneworld dynamics of the Randall-Sundrum model. 

In this respect, this connection gives a first more robust physical meaning to the generalized GUP framework that we are discussing.

The issue concerning the possibility to deal with a truncated square root GUP formulation, instead, comes out as a delicate point that can be elucidated by considering the second modified Hamilton equation, associated with the action \eqref{action_miniss}. 
It is immediate to get the dynamical equation for the momentum $p_v$:

\begin{equation} 
 \begin{split}
	\dot{p}_v =& \frac{c}{4\pi^2}\left[ \frac{3}{2}\chi p_v^2 + \frac{p_{\phi}^2}{2v^2} \right]	\sqrt{1 + 2 \beta p_v^2} \\
	=&\frac{c}{\pi^2} \frac{3}{4}\chi p_v^2\sqrt{1 + 2 \beta p_v^2}\, ,
	\label{pv_eq_GUP}
	\end{split}
\end{equation}
where, in the last equality, we have made use of the constraint (\ref{constraint_miniss}). 

\begin{figure}[ht]
	\centering
	\includegraphics[width=%
	0.48\textwidth]{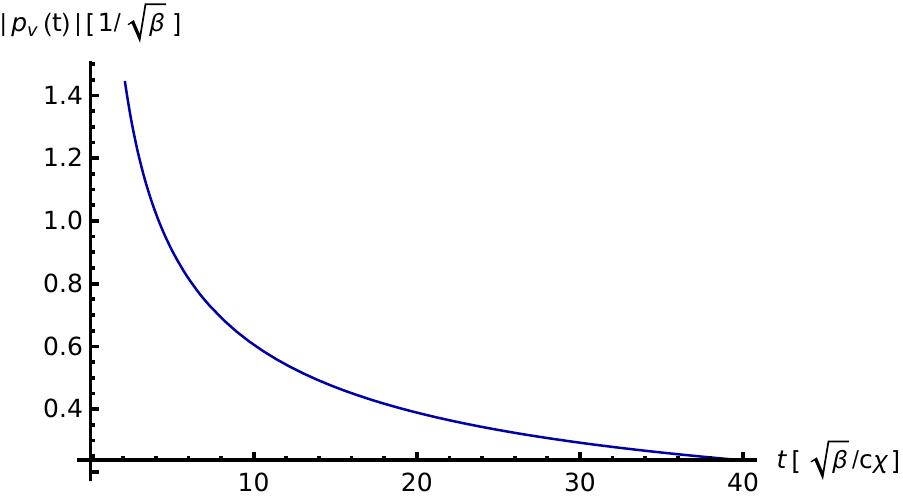}	
	\caption{Plot of the absolute value of the function $p_v(t)$, solution of the equation \eqref{pv_eq_GUP}. The conjugate momentum is naturally measured in units of $1/\sqrt{\beta}$, while the coordinate time is measured in units of $ \sqrt{\beta}  / c \chi $.
	The constant of integration is set to $k=\sqrt{2}$ in order to obtain $t=0$ as the origin of the temporal axis. }
	\label{conj_p_v}
\end{figure}

The integration of the above equation leads to a function $p_v(t)$, which, in absolute value, monotonically decreases with time and diverges towards the singularity (see Fig.~\ref{conj_p_v}).

This behavior is reflected in the Hubble parameter, which, expressed as a function of $p_v$, reads as:
\begin{equation}
    H^2 = \frac{c^2}{\pi^4}\frac{\chi^2}{16}p_v^2(1+2\beta p_v^2).
\end{equation}

From this expression we immediately see that the Hubble parameter diverges as a consequence of the presence of the singularity.

This is clearly the picture emerging from the modified classical theory, the dynamics of which is dictated by the Poisson bracket \eqref{PB_mod}, where there are no other kinematic or dynamical structures.

Now consider the following:
if we impose \emph{by hand} a constraint to the absolute value of $p_v$, the Hubble parameter will result constrained too, since it will present a maximum value.
This means that the truncation prevents the Hubble parameter to diverge, a feature that can be considered a signal of the avoidance of singularity.

This fact is not surprising \emph{per se} and what is stated has to be regarded as a heuristic observation, nevertheless, in view of this, the truncation of the momentum space in the quantum GUP formulation that we are going to discuss gains in relevance.

Indeed, this particular formulation not only is relevant in view of the emergence of a minimum length at the quantum level but also in relation to the fact that it seems to imply a cut-off physics able to affect the singularity.

Thus the truncated and non-truncated schemes appear now very different and the general picture that we have drawn seems to suggest that the Friedmann equation of the discussed model of brane cosmology is surprisingly associated, on a quantum effective level, with a zero absolute minimal uncertainty of the Universe volume. 

In conclusion, we have seen how the Minisuperspace example above well-elucidates the physical relevance of studying the truncated theory of the generalized square root GUP, as it reproduces a very different quantum dynamical paradigm with respect to the non-truncated picture. 

We also stress the fact that, in some measure, the presence of a minimal uncertainty in the truncated theory as in \cite{Kempf:1994su} legitimates the claim that this formulation only is the genuine extension of the standard GUP approach when the square root is considered. 
Clearly, the distinction between the two models (with and without the square root) becomes relevant only when the truncated momentum domain includes values of the momentum much larger than the value $\sim 1/\sqrt{\beta}$, where the expansion in series of the square root would fail.

\section{Non-truncated or full GUP theory}
  \label{section_V}
 
  We shall now follow the functional procedure we carried out in the reviewing sections.
  
  For convenience we write again the modified algebra \eqref{MM_GUP} that we are going to study:
  \begin{equation} \label{MM_GUP_2}
  	\comm{\mathbf{\hat{x}}}{\mathbf{\hat{p}}}=i \hbar \sqrt{1+2 \beta \mathbf{\hat{p}}^2}.
  \end{equation}

  The natural choice is to represent the algebra \eqref{MM_GUP_2} on momentum space.
  The position and momentum operators that we are going to construct, first of all, must be closed, densely defined and symmetric operators.
  For these reasons we define them as:
  \begin{align}
  	\mathbf{\hat{p}}\; : \quad \!\! \mathcal{D}_{\mathbf{\hat{p}}} \; &\longrightarrow  \;\mathcal{L}^2\left(\mathbb{R}, \frac{dp}{\sqrt{1+2\beta p^2}}\right)\\
  	\psi &\; \mapsto \; p \psi,
 \end{align}
 \begin{align}
  	\mathbf{\hat{x}}: \quad \mathcal{D}_{\mathbf{\hat{x}}} \; &\longrightarrow \; \mathcal{L}^2\left(\mathbb{R}, \frac{dp}{\sqrt{1+2\beta p^2}}\right)\\
  	\psi  &\;\mapsto \; i \hbar\sqrt{1+  2 \beta  p^2} \partial_{p} \psi,
  \end{align}
  where $\mathcal{D}_{\mathbf{\hat{p}}}$ and  $\mathcal{D}_{\mathbf{\hat{x}}}$ have to be dense subsets of our Hilbert space $\mathcal{H}\equiv\mathcal{L}^2\left(\mathbb{R}, \frac{dp}{\sqrt{1+2\beta p^2}}\right)$, with a deformed Lebesgue measure, introduced as usual for the symmetry of the position operator.
  For our purpose we can choose  $\mathcal{D}_{\mathbf{\hat{p}}} \equiv \mathcal{D}_{\mathbf{\hat{x}}} \equiv \mathcal{S}$, namely the Schwartz space.
  
  It is easy to prove that the chosen representation satisfies the commutator \eqref{MM_GUP_2}, even if - we again stress it - it is not the only possible one.
  As a first thing, we notice that the $\mathbf{\hat{p}}$ operator on $\mathcal{S}$ is essentially self-adjoint and that the unique self-adjoint extension is given by the adjoint of $\mathbf{\hat{p}}$ itself, which is a multiplicative operator defined on the following domain:
  	\begin{equation}
  		\begin{split}
    	\mathcal{D}_{{\mathbf{\hat{p}}^{\dagger}}}=&\left\{\psi \in \mathcal{L}^2\left(\mathbb{R},  \frac{dp}{\sqrt{1+2 \beta p^2}}\right) \mathlarger{\mathlarger{\mathlarger{\mid}}}  \right. \\
         & \quad \! \left. p \psi \in \mathcal{L}^2\left(\mathbb{R}, \frac{dp}{\sqrt{1+2 \beta p^2}}\right) \right\}.
  	    \end{split}
  \end{equation}
 Therefore from now on $\mathbf{\hat{p}}^{\dagger}:=\mathbf{\hat{p}}$ will be our "true" momentum operator.
 In this framework the momentum "eigenfunctions", in momentum representation, are Dirac deltas and their scalar product will be defined as:
 \begin{equation}
 	\braket{p}{\widetilde{p}}=\sqrt{1+2\beta p^2}\delta(p-\widetilde{p}).
 \end{equation}

 \noindent For the position operator $\mathbf{\hat{x}}$, as expected, the analysis is more subtle.
 As the explicit construction of its adjoint shows, it is not self-adjoint on $\mathcal{S}$:
  \begin{equation}
  	\begin{split}
  		\mathcal{D}_{\mathbf{\hat{x}}^{\dagger}}\!=\!& \left\{ \psi \in \mathcal{L}^2\left(\mathbb{R}, \frac{dp}{\sqrt{1+2\beta p^2}}\right) \! \mathlarger{\mathlarger{\mathlarger{\mid}}}\! \right. \\
  		& \quad \! \left. \!\exists \; \sqrt{1+2\beta p^2} \partial_{p}^{(w)} \psi \in \mathcal{L}^2\!\!\left(\mathbb{R}, \frac{dp}{\sqrt{1+2\beta p^2}}\right) \!\!\right\},
  	\end{split}
  \end{equation}
  
 thus $\mathbf{\hat{x}} \subsetneq \mathbf{\hat{x}}^{\dagger}$.

 In order to understand if this position operator $\mathbf{\hat{x}}$ is essentially self-adjoint we need to calculate the dimension of the kernel of the two operators $\mathbf{\hat{x}}^{\dagger} \pm i \mathbb{\mathbf{I}}$, a procedure that results in finding the solutions of the following differential equations:
 \begin{equation}
 	\hbar\sqrt{1+2\beta p^2}\partial_{p}^{(w)}\psi(p)= \mp \psi(p).
 \end{equation}
 By the same distributional analysis considerations that we have mentioned in Section \ref{section_II}, we can write down the solutions:
 \begin{equation}
 	\psi(p)=\kappa e^{\mp\frac{\sinh[-1](\sqrt{2\beta}p)}{\sqrt{2\beta}\hbar}}.
 \end{equation}
 Nevertheless these functions are not square-integrable functions, unless $\kappa=0$, this means that $ Ker(\mathbf{\hat{x}}^{\dagger} \pm i \mathbb{\mathbf{I}})={\mathbf{0}}$ and that $(d_+,d_-)=(0,0)$.
 Therefore, based on Von Neumann's theorem, we can conclude that, differently to the KMM case, in this framework the position operator (on $\mathcal{S}$) is essentially self-adjoint and its unique self-adjoint extension is exactly $\mathbf{\hat{x}}^{\dagger}$, defined above.
 This difference between the two formulations is non-trivial and extremely relevant.
 Indeed, as we outlined in the review section, giving up the self-adjointness of the position operator is the mathematical feature that allows the theory to host a nonzero minimal uncertainty in position, thus a "natural" minimum length.
 To prove that indeed this is the case, we now turn to the instrument provided by the DGS procedure.
 Our function $f(p)=\sqrt{1+2\beta p^2}$ goes as $f(p)\approx \abs{p}$ for $\abs{p} \gg 1$, hence the exponent $\nu$ is equal to zero.
 As pointed out in \cite{Detournay:2002fq}, in this case it is not possible to say anything about the integral function $z(p)$ a priori, but everything will depend on the precise behavior of $f(p)$.
 
 In the case under study, the quantity $z(p)$ is:
  \begin{equation}
 	z(p)=\int_{0}^{p} dq \frac{1}{\sqrt{1+2\beta q^2}}= \frac{\sinh[-1](\sqrt{2\beta}q)}{\sqrt{2 \beta}}
  \end{equation}
 and it is divergent for $\abs{p} \to \infty $.
 
 According to the DGS scheme, we are in the \textit{non-compact case}, therefore the whole procedure discussed above in the compact case, which eventually leads to finding a nonzero minimal uncertainty in position, is not available.
 
 This is still not enough to conclude that in this framework the minimal value of $\Delta{\mathbf{\hat{x}}}$ is zero.
 It is indeed necessary to prove that also in this specific case with $\nu=0$, the \textit{squeezed states} are physical states, with uncertainty in position that can be made arbitrarily small.
 As a first thing we need the general form of the squeezed states of the theory, which turns out to be:
 \begin{equation} \label{MM_sq_states}
 	\!\!\!\!\!\Phi_{\Lambda}(p)=\mathcal{N}e^{(\hbar \Lambda \eta - i \xi) \frac{\sinh[-1](\sqrt{2\beta}p)}{\hbar \sqrt{2\beta}}-\Lambda\frac{\sqrt{1+2\beta p^2}-1}{2\beta}},
 \end{equation}
  where we recall that $\eta \in \mathbb{R}$ and $\Lambda \in \mathbb{R}^{+}$.
  
\begin{figure}[ht]
 	\centering
 	\includegraphics[width=%
 	0.47\textwidth]{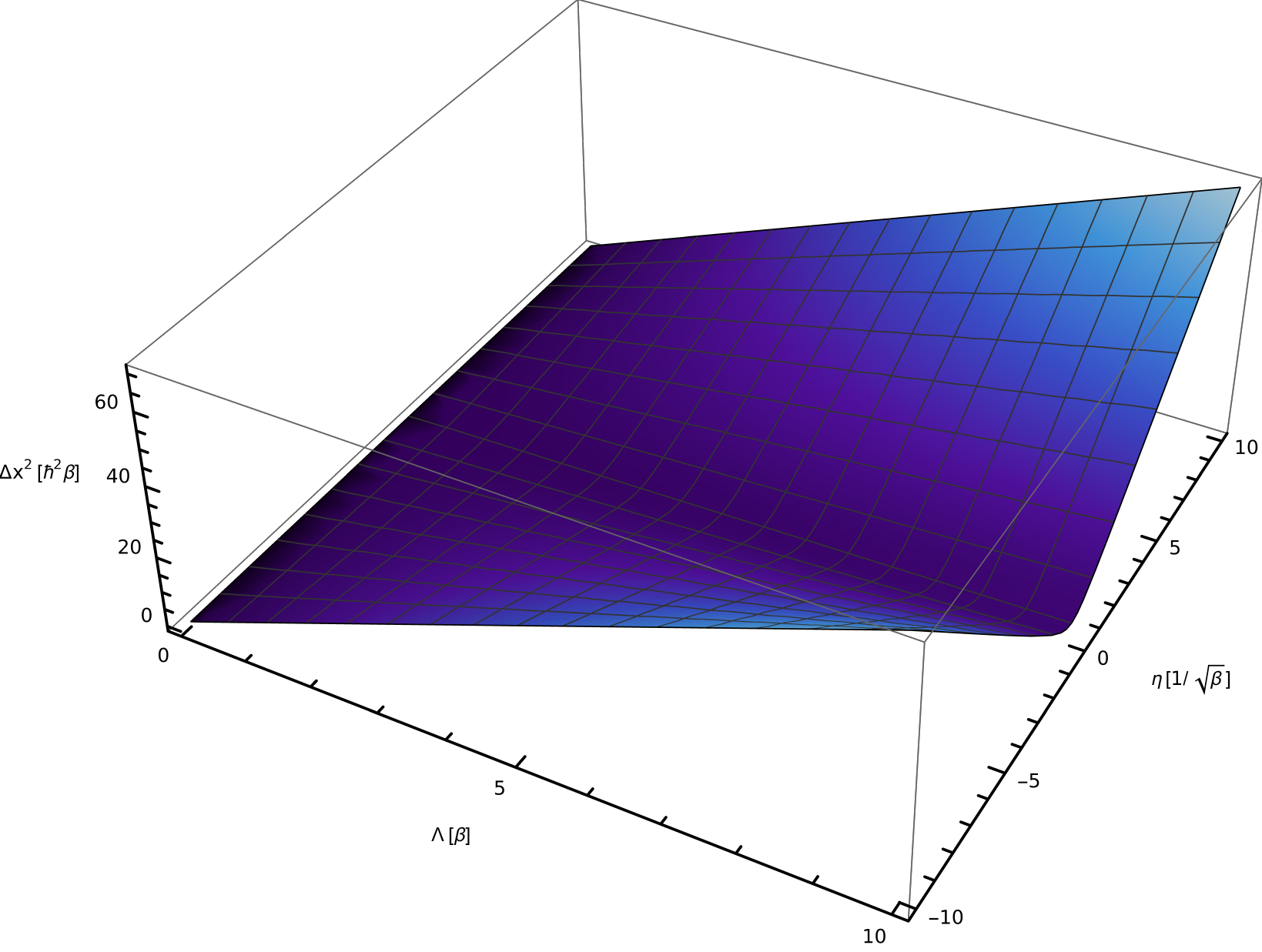}	
 		\includegraphics[width=%
 	0.47\textwidth]{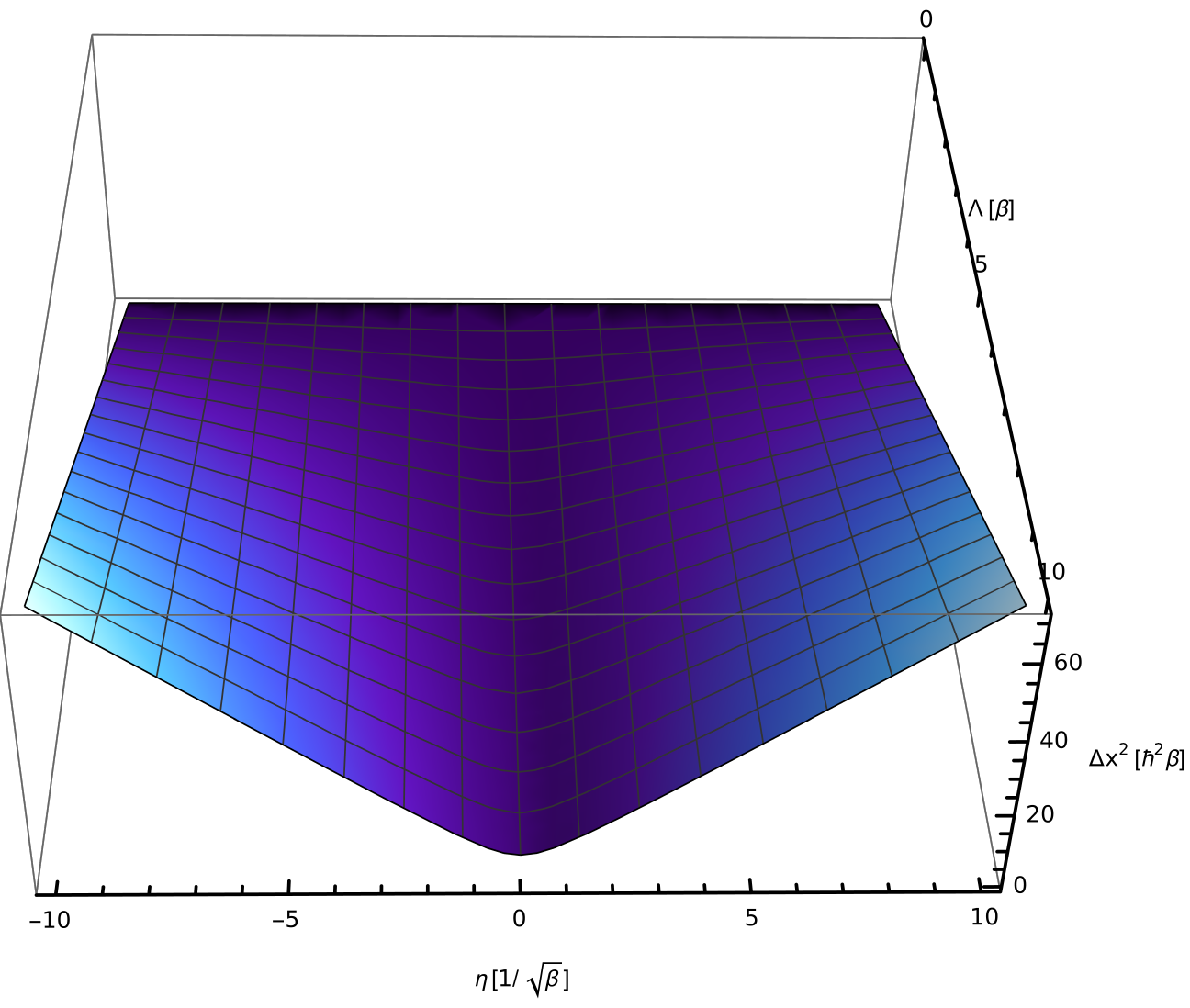}	
 	\caption{Plots of the two-dimensional surface generated by the functional $\Delta{\mathbf{\hat{x}}_{\Phi_\Lambda}}$, depending on the two real parameters $\Lambda$ and $\eta$, measured respectively in units of $\beta$ and $1/\sqrt{\beta}$, while space is measured in units of $\hbar\sqrt{\beta}$. From the two presented perspectives it is clear that the minimum of the surface with respect to $\eta$ lies along the curve $\eta=0$ and the minimum with respect to $\Lambda$ lies along the curve $\Lambda=0$.
 	Nevertheless, it is also evident how \textit{any} $\eta$-curve eventually reaches its minimum for $\Lambda=0$.}
 	\label{Lambdaeta_surface}
  \end{figure}

  The function $f(p)$  we are working with satisfies all the necessary conditions imposed by the procedure summarized in the subsection \ref{subsection_II}, as it is straightforward to verify. This allows us to use all the developed machinery to evaluate explicitly the quantity \eqref{functional_Deltax}.
  Unfortunately an analytical resolution seems not viable, thus we need to estimate numerically the value of the resulting integral as a function of the couple $(\Lambda, \eta)$.
  For a more accurate calculation we will use the following general formula to express an integral over the whole real axis as an integral over a finite interval:
  \begin{equation}
  	\int_{-\infty}^{\infty} \!\! h(x) dx \!=\! \int_{0}^{1} \!\left[ h\left(\!\frac{1}{t}\!-\!1\!\right)\!+\! h\left(\!-\frac{1}{t}\!+\!1\!\right)\right]t^{-2} dt.
  \end{equation}
 The numerical integration gives back the 2D-surface shown in Fig.~\ref{Lambdaeta_surface}, which represents the changing of the value of $\Delta{\mathbf{\hat{x}}_{\Phi_\Lambda}}$ with respect to $\Lambda$ and $\eta$, in a sample region delimited by some chosen values of the two independent variables.
 
 We are now interested in finding - if they exist at all - the values of $\Lambda$ and $\eta$ which minimizes $\Delta{\mathbf{\hat{x}}_{\Phi_\Lambda}}$.
 By visually inspecting the plot in Fig.~\ref{Lambdaeta_surface}, it appears clear that, with respect to $\eta$, the surface reaches its minimum for $\eta=0$, hence the minimum with respect to $\Lambda$ - if it exists - will lie along this specific $\eta$-curve.
 
\begin{figure}[ht]
	\centering
	\includegraphics[width=%
	0.47\textwidth]{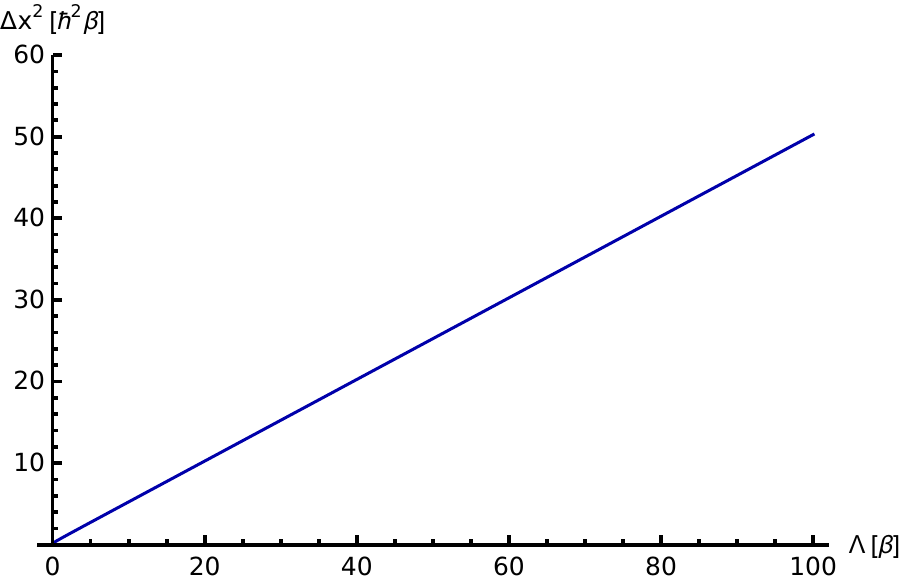}	
	\caption{Plot of the functional $\Delta{\mathbf{\hat{x}}_{\Phi_\Lambda}}$ as a function only of $\Lambda$, being $\eta=0$, in the same units of the previous plot.
	Here it can be appreciated how the minimum of the functional along the $\eta$-curve for $\eta=0$, is reached for $\Lambda=0$.
	Since physically we have that $\Lambda\neq0$, this means that the value of  $\Delta{\mathbf{\hat{x}}}$ can be made arbitrarily small for $\Lambda \to 0$ and hence that the theory does not contain a natural minimum length as nonzero minimal uncertainty in position.}
	\label{Lambdaeta_curve}
\end{figure}

 \noindent Setting $\eta=0$, we are now able to represent the changing of the value of $\Delta{\mathbf{\hat{x}}_{\Phi_\Lambda}}$ only with respect to $\Lambda$, as shown in Fig.~\ref{Lambdaeta_curve}.
 
 The graphic shows distinctly what we have anticipated: by making the value of $\Lambda$ arbitrarily small, it is possible to obtain states with an arbitrarily small uncertainty in position, asymptotically going to zero for $\Lambda \to 0$.
 Therefore we can conclude that in this theory there does not exist a nonzero minimal uncertainty in position.
 The corresponding asymptotically "maximally localized" states are hence obtained for $\eta=0$, by making the limit for $\Lambda \to 0$ of the expression \eqref{MM_sq_states}:
 \begin{equation} \label{mod_plane_waves_full_theory}
 		\Phi_{\Lambda}(p)=\frac{1}{\sqrt{2\pi\hbar}}e^{-i x \frac{\sinh[-1](\sqrt{2\beta}p)}{\hbar \sqrt{2\beta}}},
 \end{equation}
 where we have set $\mathcal{N}=1/\sqrt{2\pi \hbar}$ as usual for the normalization constant of a plane wave and $\xi=x$ because, being these states perfectly localized in position, the expectation value of $\mathbf{\hat{x}}$ coincides with the exact position $x$ of the particle itself.
 
 These states are modified plane waves and coherently they are the eigenfunctions of the position operator of the theory in momentum representation. 
 Of course they are not physical states, but they can play exactly the same role as the plane waves in the ordinary quantum theory and they can be approximated with arbitrary precision by sequences of physical states of increasing localization.
 This means that within this framework the position representation is available and it has the usual physical interpretation.
 It is immediate to write down the map from momentum space to position space:
  \begin{equation} \label{GFT_full}
 	\!\!\psi(x)\!\!:=\!\!\braket{x}{\psi}\!\!=\!\!\frac{1}{\sqrt{2\pi\hbar}}\! \int_{\mathbb{R}}\!\! \frac{dp}{\sqrt{\!1\!+\!2\beta p^2\!}} e^{i x \frac{\sinh[-1](\sqrt{2\beta}p)}{\hbar \sqrt{2\beta}}} \widetilde{\psi}(p),
  \end{equation}
  which represents a generalization of the Fourier transform, while its inverse can be written as:
  \begin{equation} \label{Inverse_GFT_full}
  \widetilde{\psi}(p):=\braket{p}{\psi}=\frac{1}{\sqrt{2\pi\hbar}} \int_{\mathbb{R}} dx \, e^{-i x \frac{\sinh[-1](\sqrt{2\beta}p)}{\hbar \sqrt{2\beta}}} \psi(x).
 \end{equation}

  For a free particle of fixed momentum $\widetilde{p}$ we have that $\widetilde{\psi}(p)=\sqrt{1+2\beta p^2}\delta(p-\widetilde{p})$ and coherently its generalized Fourier transform, through \eqref{GFT_full}, is:
  \begin{equation}
 	\psi^{\widetilde{p}}(x)=\frac{1}{\sqrt{2\pi\hbar}} e^{i x \frac{\sinh[-1](\sqrt{2\beta}\widetilde{p})}{\hbar \sqrt{2\beta}}},
  \end{equation}
 that is the same expression \eqref{mod_plane_waves_full_theory} (except for a sign), where now $p=\widetilde{p}$ is fixed and $x$ is the independent variable.
 
 By making use of the Fourier transform \eqref{GFT_full} and its inverse \eqref{Inverse_GFT_full}  it is also possible to express the action of the position and momentum operator in the x-representation:
  \begin{align}
   \bra{x}\mathbf{\hat{x}}\ket{\psi}&=x \psi(x), \\
  \begin{split}
  \bra{x}\mathbf{\hat{p}}\ket{\psi}&=\frac{1}{\sqrt{2\beta}}\sinh(-i\hbar         \sqrt{2\beta} \frac{d}{dx})\psi(x).
  \end{split}
  \end{align}
 
 \bigskip
 
 Finally we make two observations:
 \begin{itemize}
 	\item  even if we have chosen the $\eta$-curve of the surface obtained for $\eta=0$, which can be considered the most natural choice, we would have come to the same conclusion for any other value of $\eta$. Indeed, as the Fig.~\ref{Lambdaeta_surface} shows, every $\eta$-curve goes asymptotically to zero, for $\Lambda \to 0$.
    \item once we have set $\eta=0$, it is not difficult to verify through numerical integration that the resulting squeezed states are real physical states, in the sense that they belong to the domain $\mathcal{D}_{\mathbf{\hat{x}}} \cap \mathcal{D}_{\mathbf{\hat{x}}^{2}} \cap \mathcal{D}_{\mathbf{\hat{p}}} \cap \mathcal{D}_{\mathbf{\hat{p}}^2} \cap \mathcal{D}_{\comm{\mathbf{\hat{x}}}{\mathbf{\hat{p}}}}$.
\end{itemize}

 \section{Truncated or compact GUP theory}
 \label{section_VI}
 What we have just learned is that the full theory based on the algebra \eqref{MM_GUP_2} does not seem to lead to the existence of a nonzero minimal uncertainty in position, which would play the role of a "natural" minimum length.
 In light of this, it makes sense to ask whether or not some modifications of the previous framework which can account for such a desired feature are possible.
 One of the possible paths could be the implementation of the GUP theory on a one-dimensional truncated or compact momentum space.
 Besides the physical motivations given in Section \ref{section_IV}, from a more formal point of view, the arguments for such a choice are essentially two:
 \begin{itemize}
 	\item the DGS functional procedure clearly shows that whenever it is possible to recast the theory, through a proper diffeomorphism, in a compact momentum space, a nonzero minimal uncertainty in position appears.
 	Hence it should be automatic for a theory that is built on a compact momentum space \textit{ab initio} to host such a feature.
 	\item an adequate truncation of momenta could allow us to recover the series expansion method used in \cite{Fadel:2021hnx} - even if the following analysis will have to be handled differently - and it could be then compared to the DGS scheme.
 \end{itemize}
 
 Our new Hilbert space $\mathcal{H}$ will now be $\mathcal{L}^2\left( [-p_0,p_0], dp/\sqrt{1+2\beta p^2} \right)$, where $p_0$ is a generic point of the real line and we have chosen a symmetric interval with respect to zero as compact momentum space in order to preserve the obvious symmetry under parity.
 It is trivial to notice that the quantity $z(p)$ does not diverge anymore towards the endpoints of the interval:
 \begin{equation}
 	lim_{p \to \pm p_0} \frac{\sinh[-1](\sqrt{2\beta}p)}{\sqrt{2\beta}}= \pm \frac{\sinh[-1](\sqrt{2\beta}p_0)}{\sqrt{2\beta}}.
 \end{equation}
 Since we are working on a compact momentum space for construction, there is no real need to make use of the map $p \to z(p)$, which in this case is just a diffeomorphism between two compact intervals. 
 The functional analysis and all the considerations about the momentum operator $\mathbf{\hat{p}}$, the position operator $\mathbf{\hat{x}}$, their squares and the commutator are exactly the same discussed in the compact DGS case, with the only difference that now everything is written with respect to the $p$-variable, thus it will be necessary to take into account in every step the measure $\sqrt{1+2\beta p^2}$.
 
 The physical domain of the theory, therefore, will be:
 \begin{equation} \label{physical_space_CT}
 	\begin{split}
 		\mathcal{D}_{\mathbf{\hat{x}}}=&\mathlarger{\biggl\{}\psi \in \mathcal{H}^{1,2} \left([-p_0, p_0], \frac{dp}{\sqrt{1+2\beta p^2}}\right) \mathlarger{\mathlarger{\mathlarger{\mid}}} \\ &\quad \!  \psi\left(-p_0\right)=\psi\left(p_0\right)=0  \mathlarger{\biggr\}}.
 	 \end{split}\
 \end{equation}

 By applying the DGS method we are now able to find in our physical domain $\mathcal{D}_{\mathbf{\hat{x}}}$ the maximally localized states of the truncated theory and the corresponding nonzero uncertainty in position, which will be the minimum length of the theory:
 \begin{align}
 	\!\!\Psi^{\xi}(p) \!&=\!\mathcal{K}e^{-i\xi\frac{\sinh[-1](\sqrt{2\beta}p)}{\hbar \sqrt{2\beta}}}\!\!\cos(\!\frac{\pi}{2}\frac{\sinh[-1](\sqrt{2\beta}p)}{\sinh[-1](\sqrt{2\beta}p_0)}\!),  \label{max_loc_trunc}\\
 	\Delta\mathbf{\hat{x}}^{min}&=\sqrt{\frac{\beta}{2}}\frac{\pi \hbar}{\sinh[-1](\sqrt{2\beta}p_0)} \label{min_length_trunc},
 \end{align}
 where $\mathcal{K}=\sqrt{\frac{\sqrt{2\beta}}{\sinh[-1](\sqrt{2\beta}p_0)}}$.

 Of course, by construction, the state \eqref{max_loc_trunc} belongs to $\mathcal{D}_{\mathbf{\hat{x}}}$ and respects all the constraints imposed by the variational method, therefore it is a fully legitimate physical state.
 It is worth noticing that the quantity $\Delta\mathbf{\hat{x}}^{min}$ is inversely proportional, through the hyperbolic arcsine function, to $p_0$, which can be read as the half-length of the symmetric closed real interval we have chosen as momentum space.
 This implies that the larger this interval, the smaller this length will be and in the limit for $p_0 \to \infty$ we obtain $\Delta\mathbf{\hat{x}}^{min}=0$.
 
 \noindent Since the limit $p_0 \to \infty$ restores the real line as momentum space, marking the transition from the compact formulation to the non-compact one,
 this result is perfectly coherent with our conclusions about the full theory and can be interpreted as a further confirmation of what we have discussed previously.
 Furthermore, once the normalization condition is relaxed, also the maximally localized states, for $p_0 \to \infty$, are reduced to the modified plane waves \eqref{mod_plane_waves_full_theory}, which, as stated before, are the "maximally localized states" of the full theory, even if they are not proper physical states.
 
 Regarding the relation with ordinary quantum theory, it is straightforward to see that if, once we have taken the limit for $p_0 \to \infty$ - i.e. once we are dealing with the full theory - we take the limit for $\beta \to 0$, from \eqref{mod_plane_waves_full_theory},  we re-obtain the plane waves of the standard quantum mechanics as "maximally localized states", with zero uncertainty in position.
 
\noindent Furthermore, the two limits commute.
 Indeed if we first take the limit for $\beta \to 0$, the expression \eqref{max_loc_trunc} and \eqref{min_length_trunc} become:
 	\begin{align}
 	 \begin{split}
 	 &\lim_{\beta \to 0} \; \mathcal{K}e^{-i\xi\frac{\sinh[-1](\sqrt{2\beta}p)}{\hbar\sqrt{2\beta}}}\cos(\frac{\pi}{2}\frac{\sinh[-1](\sqrt{2\beta}p)}{\sinh[-1](\sqrt{2\beta}p_0)}) \\
 	& = \sqrt{\frac{1}{p_0}}e^{-i \frac{\xi}{\hbar}p}\cos(\frac{\pi}{2}\frac{p}{p_0}),
 	\end{split} \\
   	&\lim_{\beta \to 0} \sqrt{\frac{\beta}{2}}\frac{\pi \hbar}{\sinh[-1](\sqrt{2\beta}p_0)} = \frac{\pi \hbar}{2 p_0}.
   \end{align}

 These are respectively the maximally localized states and the minimal uncertainty in position, obtained through a DGS scheme, of the ordinary quantum theory implemented in a one-dimensional compact momentum space, as it can be directly verified.
 At this point, once the consistency of the first limit is accepted, by relaxing again the normalization condition, it is possible to make the limit for $p_0 \to \infty$ and once again we find the plane waves of the standard theory, with zero uncertainty in position.

 \subsection{Comparison with the series expansion procedure}
 \label{section_VIa}
  In the previous analysis, which led us to explicitly find the maximally localized states and the corresponding uncertainty in position in the truncated theory, we never specified the norm of $\mathbf{\hat{p}}$ and consequently neither the set of its possible eigenvalues, since the procedure does not require it and holds in general.
  Nevertheless, if we want to make contact with the analysis carried out in \cite{Fadel:2021hnx} and if we want to use correctly a series expansion, some conditions must be imposed.
  Indeed the series \eqref{series_exp_comm} converges to $\sqrt{1+2\beta \mathbf{\hat{p}}^2}$ if and only if $\norm{\mathbf{\hat{p}}^2}\leq 1/(2\beta)$ or equivalently if $\norm{\mathbf{\hat{p}}}\leq \sqrt{1/(2\beta)}$.
  Without loss of generality, we make the maximal choice and set  $\norm{\mathbf{\hat{p}}} = \sqrt{1/2\beta}$, meaning that the eigenvalues of $\mathbf{\hat{p}}$ belong to the set $[-\sqrt{1/(2\beta)},\sqrt{1/(2\beta)}]$, which represents our compact momentum space.
  Under these assumptions it is now possible to write:
   \begin{equation} \label{MM_gup_series_exp}
    	\Delta{\mathbf{\hat{x}}}\Delta{\mathbf{\hat{p}}} \geq \frac{\hbar}{2}\sqrt{1+ 2 \beta (\Delta{\mathbf{\hat{p}}})^2}.
   \end{equation}
  Yet, two fundamental facts must be taken into account to interpret correctly the above expression:
  \begin{itemize}
  	\item since our momentum space is a compact space, $\Delta\mathbf{\hat{p}}$ cannot take arbitrary values up to infinity, but only in the set $[0, \sqrt{2/\beta}]$;
  	\item  there are no physical states which are able to saturate the inequality, hence the equal sign must be removed.
  	This is because the squeezed states, the general form of which has been obtained in Section \ref{section_II}, have no place in the truncated formulation because they do not belong to the physical domain $\mathcal{D}_{\mathbf{\hat{x}}}$ from expression \eqref{physical_space_CT}.
  \end{itemize}
 
  At this point we can write:
   \begin{equation}
 	\Delta{\mathbf{\hat{x}}} > \frac{\hbar}{2}\sqrt{\frac{1}{\Delta{\mathbf{\hat{p}}}^2}+ 2 \beta}
   \end{equation}
  and, by inserting the value of $\Delta\mathbf{\hat{p}}$ which minimizes the left-hand side, we obtain a lower bound for 	$\Delta{\mathbf{\hat{x}}}$, namely:
   \begin{equation} \label{lower_bound_HUP}
   		\Delta{\mathbf{\hat{x}}}> \frac{1}{2}\sqrt{\frac{5}{2}}\hbar \sqrt{\beta}=  \frac{1}{2}\sqrt{\frac{5}{2}} \textit{l}_p \sqrt{\beta_0} \approx 0.79 l_{p} \sqrt{\beta_0}.
   \end{equation}
  If we now calculate the value of $\Delta{\mathbf{\hat{x}}}^{min}$ of the expression \eqref{min_length_trunc} for $p_0=\sqrt{1/(2\beta)}$ we obtain:
   \begin{equation} \label{min_unc_CT_series}
   	\begin{split}
      &\Delta{\mathbf{\hat{x}}}^{min}=\!\sqrt{\frac{\beta}{2}}\frac{\pi \hbar }{\sinh[-1](1)}\!=\! \frac{\pi}{\sqrt{2}\sinh[-1](1)} l_{p} \sqrt{\beta_0} \\
      &\approx 2.52 \, l_{p} \sqrt{\beta_0},
      \end{split}
   \end{equation}
  which is in agreement with the constraint \eqref{lower_bound_HUP}, derived from the generalized uncertainty principle \eqref{MM_gup_series_exp}.
  
  The values of $\Delta{\mathbf{\hat{x}}}$ between the lower bound just defined in \eqref{lower_bound_HUP} and the minimal uncertainty of the maximally localized states reported in \eqref{min_unc_CT_series} are evidently ruled out from the theory, since there do not exist physical states, i.e. states in $\mathcal{D}_{\mathbf{\hat{x}}}$ which respect the constraints of the DGS variational principle, which can realize them.

  \subsection{Quasi-position representation}
  \label{section_VIb}
  Even if a position representation formally still exists and can be constructed, its physical meaning in some respects is lost due to the presence of a limit in localizing physical objects.
  Indeed, while in the ordinary quantum theory the position eigenbasis, even if it is made up of non-physical states, can be approximated by a sequence of physical states of uncertainty in position decreasing to zero, this is no longer possible in our framework for the formal position eigenbasis of the $\mathbf{\hat{x}}_\lambda$ operator, hence the usual interpretation of the position representation and the density probability amplitude $\Psi(x)$ is not valid anymore. 
  Nevertheless, as pointed out first in \cite{Kempf:1994su}, information on position can still be recovered by exploiting the maximally localized states.
  In particular it is possible to project any arbitrary physical state $\ket{\psi}$ onto the maximally localized state $\ket{\xi}$, defining in this way the probability amplitude of finding the particle maximally localized around the position $\xi$.
  In this way the maximally localized states of the theory can be interpreted as constituting a basis for a new representation, namely the \textit{quasi-position representation}: 
  \begin{equation} \label{GFT_truncated}
  	\begin{split}
  		\psi(\xi)\!:=\!\braket{\xi}{\psi}\!\!=\!\!&\int_{-p_0}^{p_0}  \!\! \frac{d{p}}{\sqrt{1+2\beta p^2}} \biggl\{\mathcal{K}e^{
  			i \xi \frac{\sinh[-1](\sqrt{2\beta}p)}{\sqrt{2\beta} \hbar}} \\ & \! \times \! \cos(\frac{\pi}{2}\frac{\sinh[-1](\sqrt{2\beta}p)}{\sinh[-1](\sqrt{2\beta}p_0)})\widetilde{\psi}(p)  \! \biggr\}. 
  	\end{split}
  \end{equation}

\begin{figure}[htbp]
  	\centering
  	\includegraphics[width=%
  	0.47\textwidth]{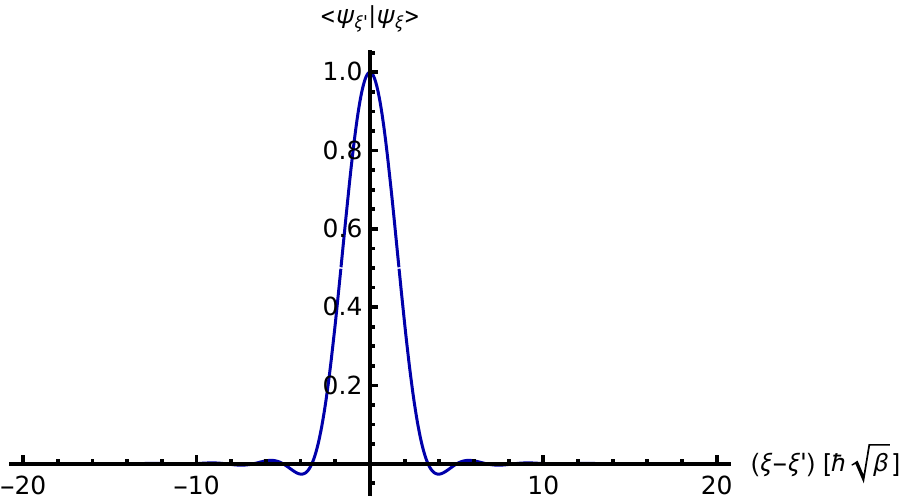}	
  	\caption{Plot of the scalar product between maximally localized states as function of the difference $(\xi-\xi')$ in units of $\hbar\sqrt{\beta}$, for the arbitrary choice of the real closed interval of momentum space $[-p_0,p_0]=[-5,5]$, where the momentum is measured in units of $1/\sqrt{\beta}$.}
  	\label{ps_max_f}
  \end{figure}

  These wave functions are consequently called \textit{quasi-position wave functions}.
  We notice that the basis made by the maximally localized states is not orthogonal (see Fig.~\ref{ps_max_f}):
  \begin{equation}
  \begin{split}
    	&\braket{\psi_{\xi'}^{ml}}{\psi_{\xi}^{ml}}\!\!=\!\!\!\int_{-p_0}^{p_0} \!\!\! \frac{d{p}}{\sqrt{1+2\beta p^2}} \left\{\mathcal{K}^2e^{i (\xi-\xi') \frac{\sinh[-1](\sqrt{2\beta}p)}{\sqrt{2\beta} \hbar}} \right. \\
  		&  \qquad \qquad \quad  \times \!\left. \cos[2](\frac{\pi}{2}\frac{\sinh[-1](\sqrt{2\beta}p)}{\sinh[-1](\sqrt{2\beta}p_0)})\right\}\\
		 &=-2 \sqrt{2} \pi ^2  \beta ^{3/2} \hbar^3 \sin \left(\frac{(\xi'-\xi ) \sinh ^{-1}\left(\sqrt{2\beta}   {p_0}\right)}{\sqrt{2 \beta} \hbar}\right)  \\
		 &\quad \times \bigg[2 \pi ^2 \beta  \hbar^2 (\xi - \xi') \sinh ^{-1}\left(\sqrt{2 \beta} {p_0}\right)\\
		 &\qquad \quad  +4(\xi'-\xi )^3 \sinh ^{-1}\left(\sqrt{2\beta} {p_0}\right)^3\bigg]^{-1}.
	\end{split}
  \end{equation}
  This is analog to what happens in the original GUP formulation in \cite{Kempf:1994su}, where the lack of the orthogonality property of the quasi-position basis is attributed to the "fuzziness" of the space.
  
  The map \eqref{GFT_truncated} from momentum space to quasi-position space is clearly a generalization of the Fourier transform.
  In order to see that this object is well defined, expression \eqref{GFT_truncated} can be rewritten as:
   	\begin{align} \label{GFT_truncated_PW}
     &\psi(\xi)=\int_{-q_0}^{q_0} dq \sqrt{\frac{1}{q_0}}e^{i \xi \frac{q}{\hbar}} \widetilde{\psi}(q) \cos(\frac{\pi}{2}\frac{q}{q_0}), \\
  	&\text{where} \quad  q:=\frac{\sinh[-1](\sqrt{2\beta}p)}{\sqrt{2\beta}}.
  	\end{align}
  What we have obtained now is a standard Fourier transform of the compactly supported function $\widetilde{\psi}(q)\cos(\frac{\pi}{2}\frac{q}{q_0})$.
  
  \noindent By momentarily promoting $\xi$ to be a complex variable, we can make use of the Paley-Wiener theorem that assures us that the function $\psi(\xi)$ exists and in particular that it is an entire complex function, which is square-integrable over horizontal lines in the complex plane and therefore also for real values of $\xi$, the only ones in which we are interested, being $\xi$ the position expectation value of an arbitrary state.
  It is interesting to notice that, since we are dealing with compactly supported functions, it is possible to express $\psi(\xi)$ as a power series of $\xi$:
   \begin{equation}
 	\begin{aligned}
 		\psi(\xi)&=\sum_{n=0}^{\infty} a_{n}\xi^{n}, \\
 		a_{n}&=\frac{1}{n!} \int_{-q_0}^{q_0} dq \sqrt{\frac{1}{q_0}} \phi(q) \cos(\frac{\pi}{2}\frac{q}{q_0}) \left(\frac{i q}{\hbar}\right).
 	\end{aligned}
   \end{equation}
  It can be shown that this series is absolutely convergent and this implies that the $\psi(\xi)$ functions are $C^{\infty}$-smooth, as they should.
  If we now choose as $\widetilde{\psi}(p)$ the momentum "eigenfunction" $\sqrt{1+2\beta p^2}\delta(p-\widetilde{p})$, through the map \eqref{GFT_truncated} we obtain:
  \begin{equation} \label{free_particle_quasi_pos}
  	\psi^{\widetilde{p}}(\xi)=\mathcal{K}\cos(\frac{\pi}{2}\frac{\sinh[-1](\sqrt{2\beta}\widetilde{p})}{\sinh[-1](\sqrt{2\beta}p_0)}) e^{i \xi \frac{\sinh[-1](\sqrt{2\beta}\widetilde{p})}{\sqrt{2\beta} \hbar}}.
  \end{equation}

  The function \eqref{free_particle_quasi_pos}, which is a modified plane wave, represents of course a free particle in quasi-position representation, with fixed momentum $\widetilde{p}$ and fully delocalized in the $\xi$-space.
  The obtained Fourier map is invertible and the inverse transform can be obtained by starting from \eqref{GFT_truncated_PW}:
  \begin{equation} \label{InverseGFT_trunc}
  	\widetilde{\psi}(q)\Theta^{q+q_0}_{-q+q_0}=\frac{1}{2\pi\hbar}\frac{\sqrt{q_0}}{\cos(\frac{\pi}{2}\frac{q}{q_0})} \int_{-\infty}^{\infty} \! \! \! d\xi \psi(\xi) e^{-i \xi q / \hbar},
  \end{equation}
  where $\Theta^{q+q_0}_{-q+q_0}:=\Theta(q+q_0) \times \Theta(-q+q_0)$ is the product of two Heaviside functions, the natural presence of which signals that the inverse map is correctly giving back compactly supported functions in the interval $[-q_0,q_0]$.
  By making use of the relation $q(p)$ we can of course come back to the $p$ variable and rewrite \eqref{InverseGFT_trunc} as a function of $p$.
  
  It is natural at this point to ask which is the action of the momentum operator and position operator in the quasi-position representation. 
  By carefully using the definition of the generalized Fourier transform \eqref{GFT_truncated} it is possible to show what follows:
  \begin{align}
  \bra{\xi}\mathbf{\hat{p}}\ket{\psi}&=\frac{1}{\sqrt{2\beta}}\sinh(-i\hbar         \sqrt{2\beta} \frac{d}{d\xi})\psi(\xi), \\
  \begin{split}
  \bra{\xi}\mathbf{\hat{x}}\ket{\psi}&=\xi \psi(\xi) +              \frac{\pi}{2}\frac{i\hbar\sqrt{2\beta}}{\sinh[-1](\sqrt{2\beta}p_0)} \\
  &\quad \times \tan(-\frac{\pi}{2}\frac{i\hbar\sqrt{2\beta}}{\sinh[-1](\sqrt{2\beta}p_0)} \frac{d}{d\xi})\psi(\xi).
  \end{split}
  \end{align}

  As expected, they are non-local differential operators and their action can be made explicit by a series expansion in the derivative operator itself. Nevertheless for a generic function $\psi(\xi)$ the series is in general not convergent. 
  
  \section{Wave Packets}
  
  \label{section_VII}
  We want now to explore some physical consequences of the theory within both formulations by studying one of the simplest physical systems, that is a free wave packet.
  Exactly as in the ordinary quantum theory, we can construct a wave packet evolving in time as a superposition of time-dependent plane waves:
  \begin{equation}
  	\Psi (x,t) =\frac{1}{\sqrt{2\pi\hbar}} \int_{-\infty}^{+\infty} \widetilde{\Psi}(p)e^{\frac{i p x}{\hbar}-\frac{i t E(p)} {\hbar}} dp,
  \end{equation}
  where $\widetilde{\Psi}(p)$ is the ordinary Fourier transform of the function $\Psi (x,t)$ at $t=0$ and $E(p)$ is the dispersion relation between energy and momentum.
  On this ground, in the full GUP theory we will use as infinite basis for the wave packet the modified plane waves \eqref{mod_plane_waves_full_theory}, which correctly are the eigenfunctions of the position operator in momentum representation:
   \begin{equation}
   	\begin{split}
 	\Xi(x,t)=\frac{1}{\sqrt{2\pi\hbar}}&\int_{-\infty}^{+\infty}  \frac{dp}{\sqrt{1+2\beta p^2}} \biggl\{\widetilde{\Xi}(p) \\
 	& \times e^{\frac{i x \sinh[-1](\sqrt{2\beta}p)}{\sqrt{2\beta}\hbar}-\frac{i t  E(p)}{ \hbar}}\biggr\},
 	\end{split}
    \end{equation}
  where $\widetilde{\Xi}(p)$ is obtained via the generalized Fourier transform \eqref{GFT_full} of $\Xi(x,t)$ at $t=0$.
  On the other hand, in the compact theory, to recover physical information on position we need to rely on quasi-position representation. It is therefore natural to use maximally localized states as infinite basis for the construction of the wave packet.
  Coherently we notice that in the ordinary quantum theory and in the full GUP theory the plane waves and the modified plane waves used as basis for the wave packet can be obtained respectively as the Fourier transform and generalized Fourier transform of a Dirac delta $\delta(p-\widetilde{p})$.
  This holds true in the compact theory as well, where via the generalized Fourier transform \eqref{GFT_truncated} of a Dirac delta $\delta(p-\widetilde{p})$ we obtain the states \eqref{free_particle_quasi_pos}.
  In light of this we can write:
   \begin{equation}
   	\begin{split}
 		\Phi(\xi,t)=&\int_{-p_0}^{+p_0} \frac{dp}{\sqrt{1+2\beta p^2}} \biggl\{\widetilde{\Phi}(p) e^{\frac{i \xi \sinh[-1](\sqrt{2\beta}p)}{\sqrt{2\beta}\hbar}-\frac{i t  E(p)}{\hbar}}\\ 
 		&\times \cos(\frac{\pi}{2}\frac{\sinh[-1](\sqrt{2\beta}p)}{\sinh[-1](\sqrt{2\beta}p_0)})\biggr\}.
 	\end{split}
    \end{equation}
   
   \bigskip
   
  Since we are interested in free motion, the dispersion relation in both the GUP theories will be $E(p)=p^2/2m$, where $m$ is the mass of the particle.
  
   \begin{figure}[ht]
  	\centering
  	\includegraphics[width=%
  	0.47\textwidth]{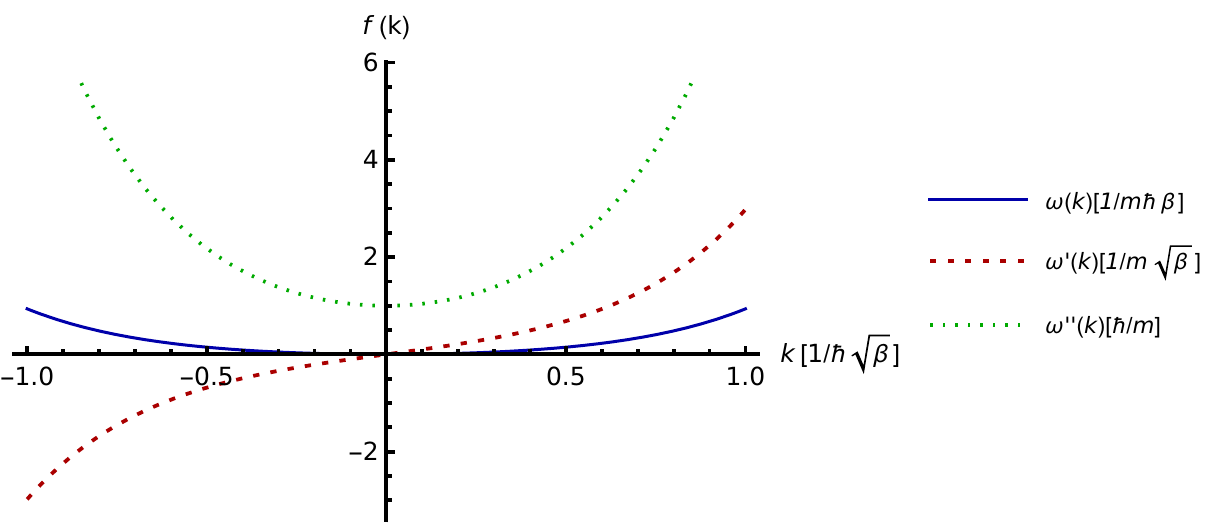}	
  	\caption{Plot of the GUP-modified dispersion relation \eqref{GUP_rel_disp} for a free particle together with its first derivative, which represents the group velocity of the wave packet and its second derivative, which instead is responsible for the dispersion of the wave packet.}
  	\label{rel_disp_GUP}
  \end{figure}
  
  \noindent This is the same one as the ordinary theory since the free particle Hamiltonian 
  is left untouched by the modification of the commutator.
  Nevertheless, if we express the dispersion relation in terms of the frequency $\omega$ and the wave number $k$ we are able to appreciate the deep difference between the GUP theories and the standard one (see Fig.~\ref{rel_disp_GUP}):
 	\begin{flalign}
  		&\omega(k)=\frac{\hbar k^2}{2 m} \qquad \qquad \qquad  \!\! \text{standard theory} \label{standard_rel_disp},\\
  		&\omega(k)=\frac{\sinh[2](\sqrt{2\beta}\hbar k)}{4m \hbar \beta} \qquad  \text{GUP theories} \label{GUP_rel_disp}.
  	\end{flalign}
   
  Here it is important to pay attention to the fact that while in the full GUP theory the dispersion  relation \eqref{GUP_rel_disp} does not contain boundaries on the possible values of $k$ and $\omega$, in the compact theory, since the momentum is constrained in some interval $[-p_0, p_0]$, $k$ will be automatically limited and this leads to the existence of a minimum wavelength $\lambda$, as it would be expected from the presence of a minimum length in the theory. 
  It is also interesting to notice that the compactness of the momentum space implies an upper bound on the angular frequency $\omega$, which could be interpreted as a lower bound on the possible time interval. 
  In order to make a comparison between the time-evolution of a wave packet in the three different frameworks we will analyze wave packets built up by fixing a gaussian-like wave function in momentum space, being careful that the chosen states belong to the physical domain of the different theories.
  
 \bigskip
  
  \onecolumngrid
   
  \bigskip
  
  \noindent Formally, we will have:
  
  \begin{align}
  	\begin{split}
  		&\text{Standard theory}: \\
  		&\Psi (x,t) =\frac{1}{\sqrt{2\pi\hbar}} \int_{-\infty}^{+\infty} dp \;  \mathcal{A}e^{-\frac{(p-\gamma)^2}{2 \sigma_{p}}}e^{i \frac{p x} {\hbar}-i t \frac{p^2}{2m \hbar}}     \label{Gaussian_ST},
  	\end{split} \\
     \begin{split}
    	& \text{Full GUP theory}: \\
    	& \Xi(x,t)=\frac{1}{\sqrt{2\pi\hbar}}\int_{-\infty}^{+\infty} \frac{dp}{\sqrt{1+2\beta p^2}}\mathcal{B}e^{-\beta \frac{(p-\nu)^2}{2\sigma_{p}}} e^{i x \frac{\sinh[-1](\sqrt{2\beta}p)}{\sqrt{2\beta}\hbar}-i t \frac{p^2}{2m \hbar}} \label{Gaussian_FT},
    \end{split} \\
    \begin{split}
    	& \text{Compact GUP theory}:\\
    	&\Phi(\xi,t)\!\!=\!\!\int_{-p_0}^{+p_0} \!\!\!\! \frac{dp}{\sqrt{1+2\beta p^2}} \mathcal{C}e^{-\beta \frac{(p-\kappa)^2}{2\sigma_{p}}}\cos[2](\frac{\pi}{2}\frac{\sinh[-1](\sqrt{2\beta}p)}{\sinh[-1](\sqrt{2\beta}p_0)}) e^{i \xi \frac{\sinh[-1](\sqrt{2\beta}p)}{\sqrt{2\beta}\hbar}-i t \frac{p^2}{2m\hbar}} \label{Gaussian_CT}.
    \end{split}
  \end{align}
  where $\mathcal{A}, \mathcal{B}, \mathcal{C}$ are the normalization constants, $\gamma, \nu, \kappa$ are real parameters and $\sigma_p$ is a real positive parameter.

    \medskip
  
    \begin{figure*}[htbp]
  	\centering
  	\includegraphics[width=%
  	0.47\textwidth]{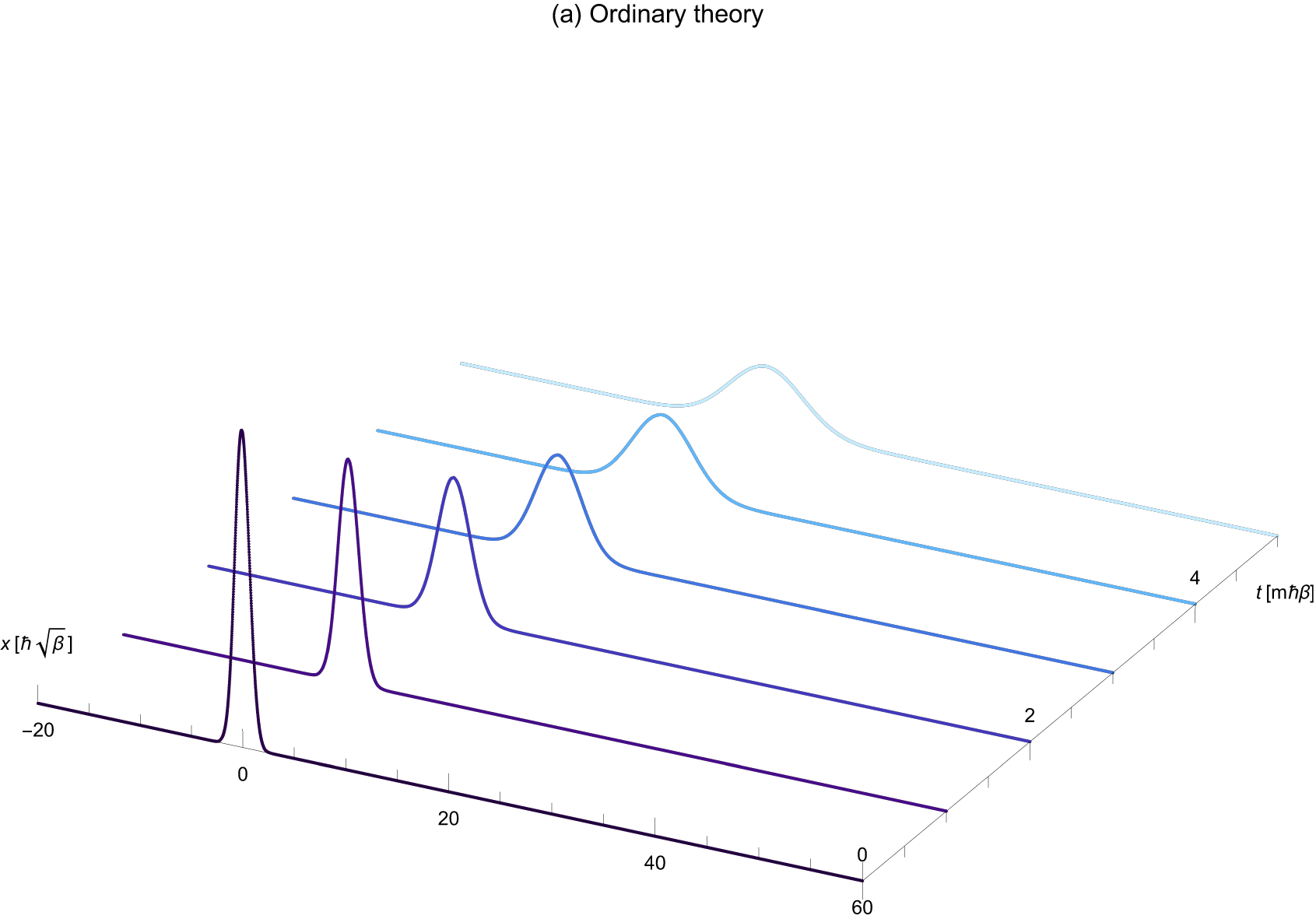}
  	\includegraphics[width=%
  	0.47\textwidth]{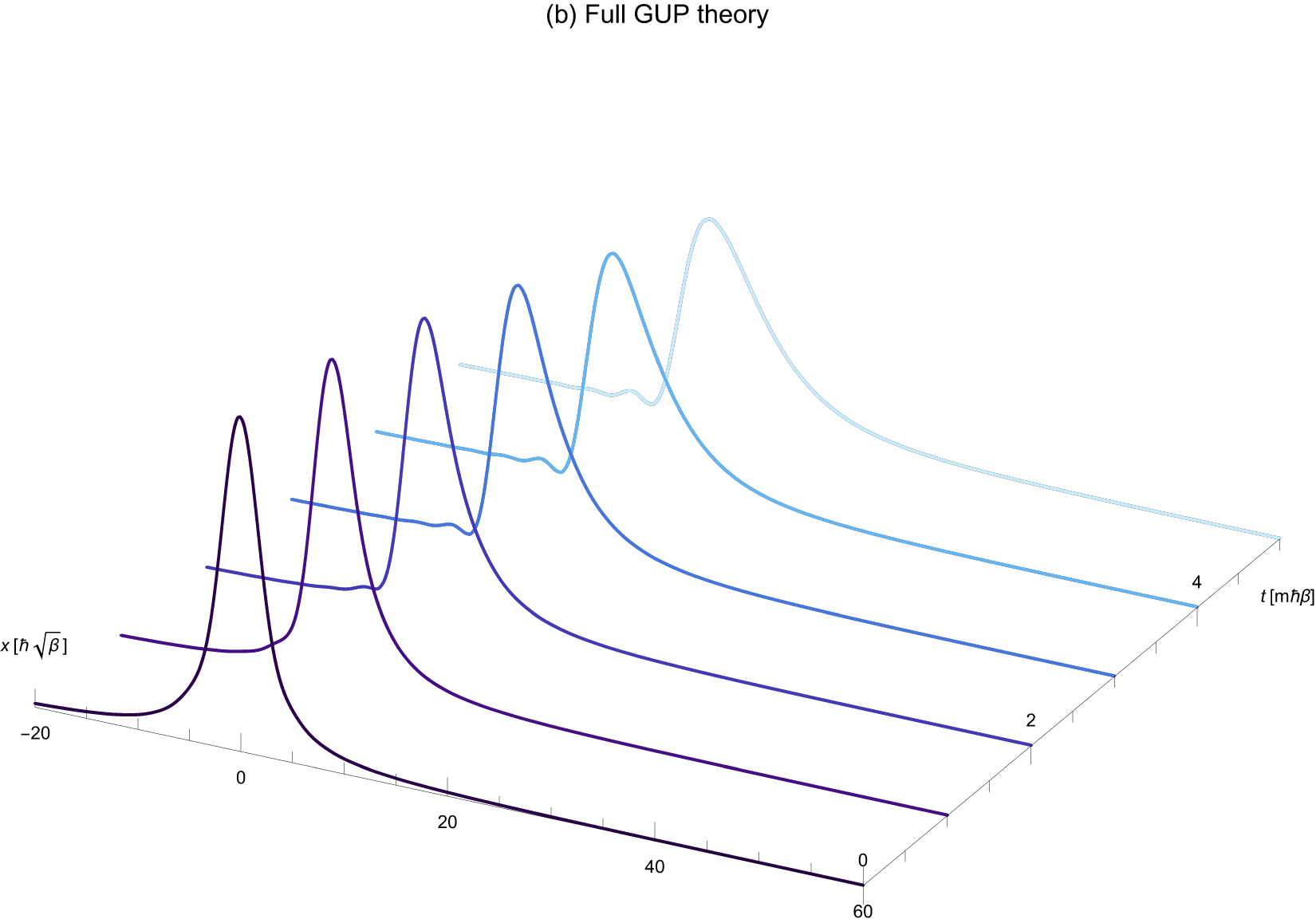}
  	
    \bigskip
    \bigskip
    
  	\includegraphics[width=%
  	0.47\textwidth]{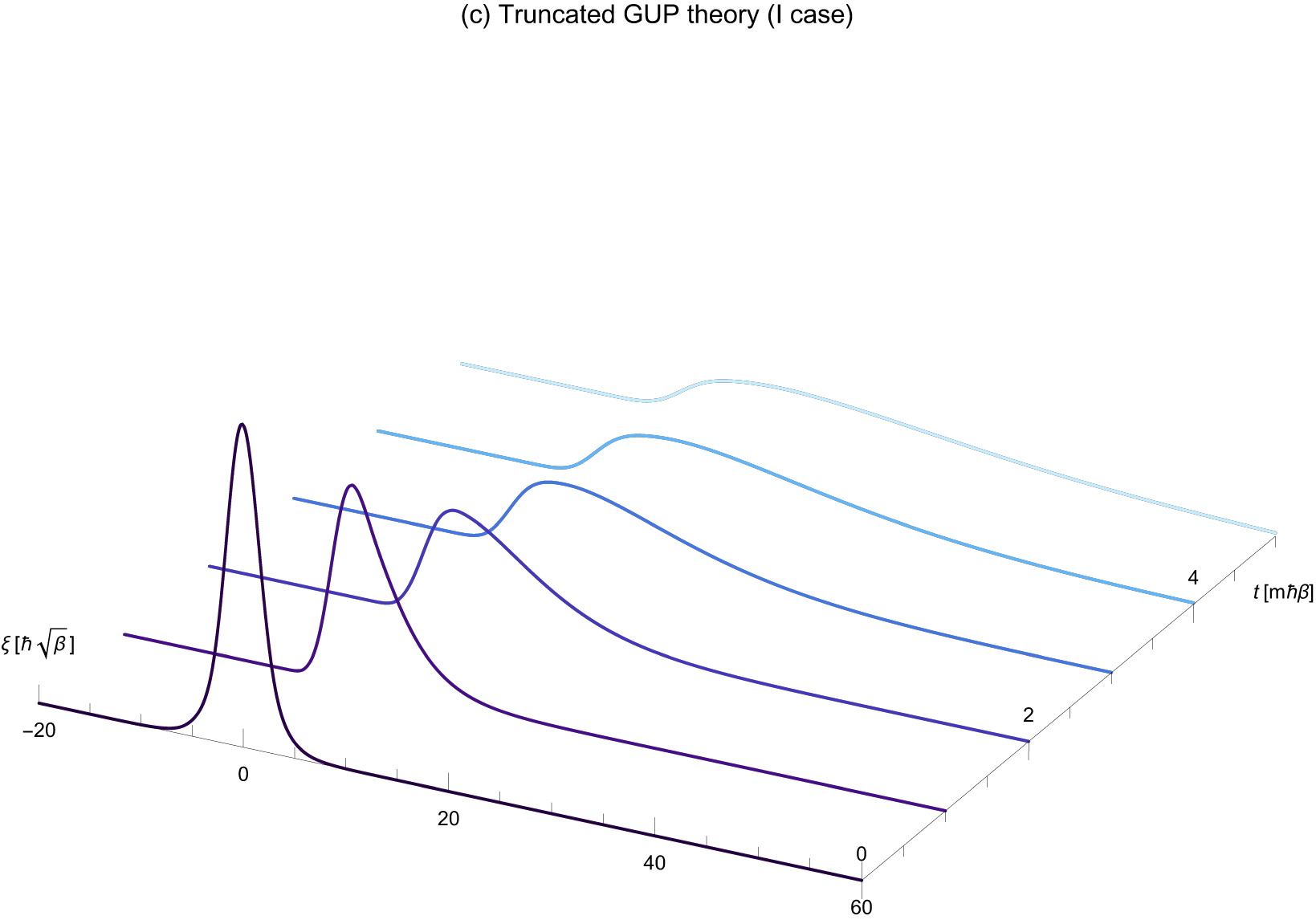}
  	\includegraphics[width=%
  	0.47\textwidth]{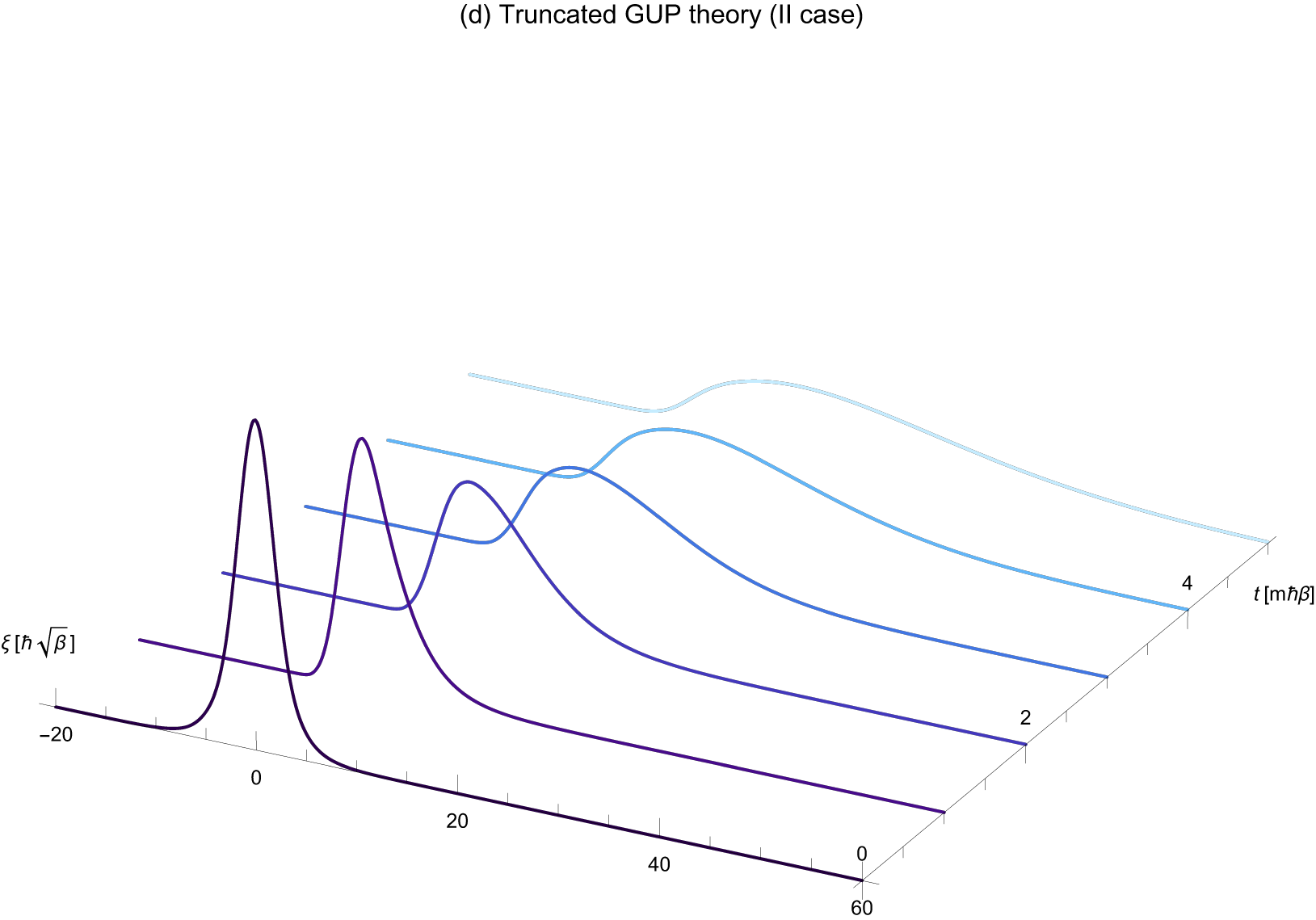}	
  	\caption{Plots of the spreading in time of the squared modulus of the wave packets \eqref{Gaussian_ST}-\eqref{Gaussian_CT} in the different frameworks here discussed.
  	Space is naturally measured in units of $\hbar\sqrt{\beta}$, time in units of $m\hbar\beta$ and momentum in units of $1/\sqrt{\beta}$.
  	The parameter $\sigma_{p}$ is set equal to unity in all the wave packets.
  	In figure (a) is shown the spreading of the gaussian wave packet \eqref{Gaussian_ST} in the ordinary quantum theory, in figure (b) is possible to appreciate the spreading of the wave packet \eqref{Gaussian_FT} in the full GUP theory, while in figures (c) and (d) is exhibited the spreading of the wave packet \eqref{Gaussian_CT} in the compact GUP theory for two different choices of the closed interval of the momentum space, respectively $[-p_0,p_0]=[-5,5]$ and $[-p_0,p_0]=[-3.5,3.5]$, in units of $1/\sqrt{\beta}$. }
  	\label{wp_spread}
  \end{figure*}
  
  \twocolumngrid
  
  \bigskip
  
  The quantity $\gamma$ in \eqref{Gaussian_ST} represents, in the ordinary theory, the (initial) expectation value of the momentum operator $\mathbf{\hat{p}}$ for the considered state, but this is not true for the parameters $\nu$ and $\kappa$ in \eqref{Gaussian_FT} and \eqref{Gaussian_CT} in the GUP theories.
  Since we want to compare states with the same initial conditions the parameters $\nu$ and $\kappa$ will hence be fixed in order to have $\expval{\hat{\mathbf{p}}}=\gamma$ also for the wave packets in the two modified theories.
  On the other hand, the initial $(t=0)$ expectation value $\expval{\hat{\mathbf{x}}}$ for the position operator is automatically zero for all the wave packets.
  
  \noindent Numerical evaluation of these integrals - for which analytical solutions seem not available - are shown in the graphics below in Fig.~\ref{wp_spread}, where the probability density at different times for the three wave packets is plotted, for an arbitrary yet proper choice of the free parameters.
  
\begin{figure}[htbp]
 	\centering
 	\includegraphics[width=%
 	0.47\textwidth]{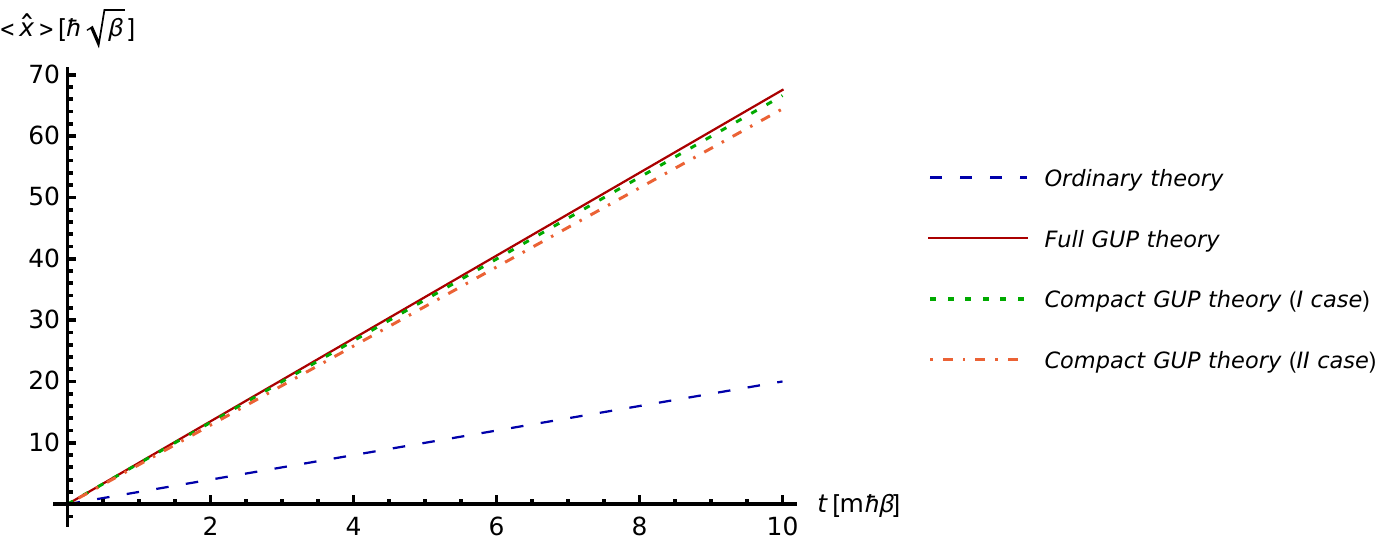}	
 	\caption{Plot of the expectation value of the position operator in units of $\hbar \sqrt{\beta}$ as a function of time, measured in units of $m\hbar\beta$, for the different wave packets in the three quantum frameworks here considered.
 		We can notice how the difference in the relations \eqref{exp_val_H_picture1}-\eqref{exp_val_H_picture2} here results in wave packets with an expectation value of the position changing more rapidly in the GUP theories with respect to the ordinary quantum theory.}
 	\label{exp_val_pos_wp}
 \end{figure}

\begin{figure}[htbp]
  	\centering
  	\includegraphics[width=%
  	0.47\textwidth]{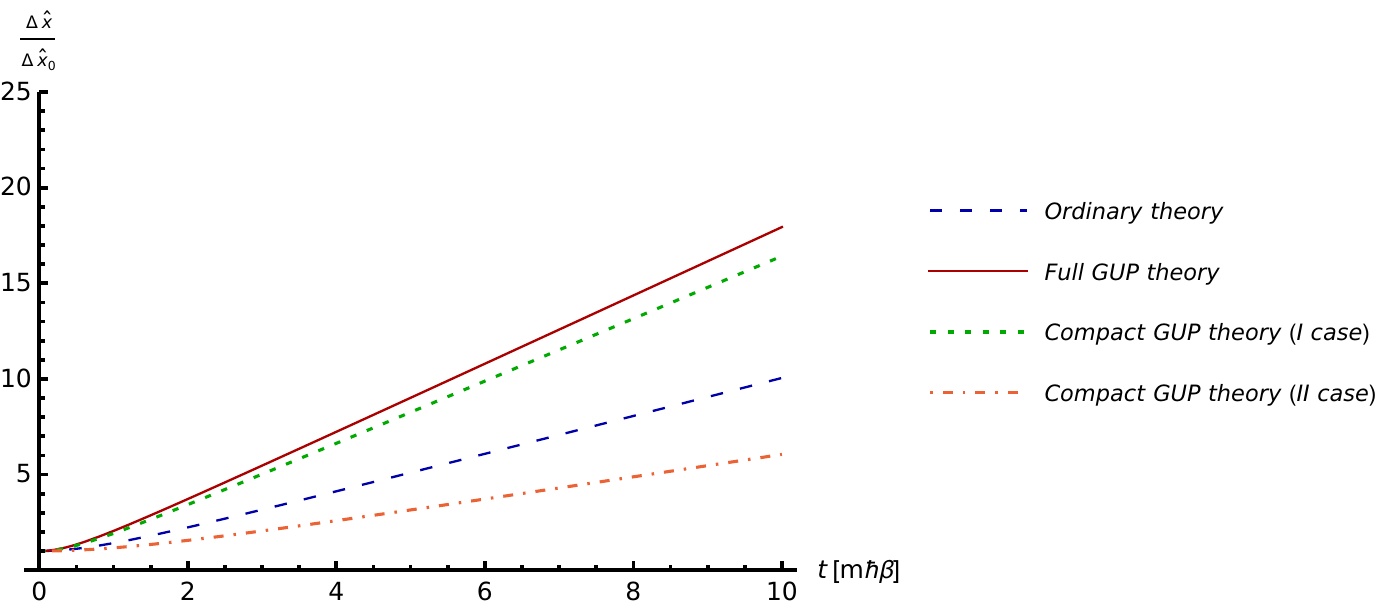}	
  	\caption{Plot of relative uncertainty in position as a function of time for the different wave packets studied in the three quantum frameworks, in the same units of the previous plot.
  		We are able to see how the wave packet spread in the full GUP theory is always more rapid with respect to the one in the ordinary theory, while the wave packet spread in the compact GUP theory strongly depends on the real interval chosen as a momentum space, producing thus physical objects which can spread more or less rapidly with respect to the one in the standard theory.}
  	\label{rel_wp_spread}
  \end{figure}

 A more quantitative picture of the situation can be obtained by inspecting the plot in Fig.~\ref{exp_val_pos_wp} of the expectation value of position $\expval{\hat{\mathbf{x}}}$ and the plot in Fig.~\ref{rel_wp_spread} of the relative uncertainty in position $\Delta\mathbf{\hat{x}}/\Delta\mathbf{\hat{x}}_0$ as a function of time for the three cases, where $\Delta\mathbf{\hat{x}}_0$ is the initial uncertainty.
 
 From the first plot we notice that, even if $\expval{\hat{\mathbf{p}}}$ is the same for all the wave packets, the time-evolution law for $\expval{\hat{\mathbf{x}}}$ is different.
 This can be easily understood by looking at the evolution of the $\hat{\mathbf{x}}$ operator in the Heisenberg picture in the different frameworks:
  \begin{align} 
  	\frac{d\hat{\mathbf{x}}}{dt}\!=\! \frac{i}{\hbar}\comm{\frac{\hat{\mathbf{p}}^2}{2m}}{\hat{\mathbf{x}}^2} \!\!\Rightarrow \!\hat{\mathbf{x}}(t)\!=&\hat{\mathbf{x}}(0)+ \frac{\hat{\mathbf{p}}}{m} t, \label{exp_val_H_picture1}\\
  	\frac{d\hat{\mathbf{x}}}{dt}\!=\! \frac{i}{\hbar}\comm{\frac{\hat{\mathbf{p}}^2}{2m}}{\hat{\mathbf{x}}^2} \!\!\Rightarrow \!\hat{\mathbf{x}}(t)\!=&\hat{\mathbf{x}}(0)\!+ \!\sqrt{1+2\beta\hat{\mathbf{p}}^2} \frac{\hat{\mathbf{p}}}{m} t. \label{exp_val_H_picture2}
  \end{align}

  From here it is clear why the expectation values of the position operator are different. 
  By looking at the second plot in Fig.~\ref{rel_wp_spread}, instead, it is evident how differently the wave packets spread in the different frameworks.
   
  In particular we see that the wave packet of the full GUP theory spreads more rapidly than the wave packet of the ordinary theory, while the spreading of the wave packet in the truncated GUP theory really depends on how the compact interval of momentum is fixed. 
  Thus, according to the chosen interval, we can have wave packets spreading more or less rapidly with respect to the ordinary theory, but always more slowly than the full GUP theory, the spreading curve of which represents an upper limit for the region that the position uncertainty of these wave packets can explore.

  We stress the fact that within the compact GUP theory is then possible to obtain wave packets that spread really slowly in time and in the end this is due to the truncation process which cuts out all the modified plane waves with higher momentum.
  
  \section{Concluding remarks}
  \label{section_VIII}
  We have analyzed in detail, in one dimension, the extended formulation of the GUP theory deriving from a square root-modified Heisenberg algebra, which verifies the Jacobi identity, as proposed in \cite{Fadel:2021hnx}.
  
  The main merit of our study is to have proved that, differently from what was stated in \cite{Fadel:2021hnx} and differently from the original approach 
  in \cite{Kempf:1994su}, 
  the considered formulation, without any truncation of the momentum space, is not associated with a minimal uncertainty different from zero for the position operator.
  This result was first of all signaled by the functional analysis of the position operator, which has resulted to be essentially self-adjoint, exactly as in the ordinary quantum theory
  and differently from the KMM GUP theory, and it was then supported by considerations regarding the modified Lebesgue measure of the theory, the integral of which is divergent, and 
  consequently by an explicit calculation carried out according to the functional methods presented in \cite{Detournay:2002fq}.
   To obtain a minimal uncertainty for the position operator different from zero and, in this sense, to extend the original formulation, we have shown that a truncation (by hand) of the momentum space is necessary. Then, we constructed the so-called quasi-position representation and, by following a similar scheme to that one presented in \cite{Kempf:1994su}, we arrived 
  at a complete characterization of the modified quantum theory.
  From a physical point of view, a significant difference with respect to the original analysis consists in having obtained a minimal uncertainty in position 
  realized by states that do not belong to the boundary of the uncertainty relation (i.e. when the equality sign holds).
  These states are indeed ruled out from the theory since they cannot satisfy the boundary conditions of the obtained physical domain \eqref{physical_space_CT}.
  
  \noindent In particular, whenever the truncation is chosen in such a way that is possible to apply the series expansion method discussed in \cite{Fadel:2021hnx}, 
  the uncertainty relation, which can be explicitly found in this case, is strictly an inequality, setting a lower bound $\Delta\mathbf{\hat{x}}^{GUP}$ for the value of $\Delta\mathbf{\hat{x}}$, compatible with the value of $\Delta\mathbf{\hat{x}}^{min}$ obtained in our analysis through the methods discussed above. 
  In this specific case, this fact suggests that all the states which have minimal uncertainty in position in the range $]\Delta\mathbf{\hat{x}}^{GUP},\Delta\mathbf{\hat{x}}^{min}[$ must correspond to non-physical states as well, for which, for instance, the energy is diverging or not well-defined. 
  Finally, we have analyzed the spreading of localized wave packets both in the truncated and non-truncated theory, comparing the obtained properties with the standard ones coming from the ordinary
  quantum mechanics. 
  We have shown that, at equivalent initial conditions, the non-truncated or full theory displays wave packets that spread more rapidly than the ordinary quantum theory. Instead, 
  the truncated or compact theory exhibits spreading features faster or slower than ordinary quantum mechanics, depending on the width of the real closed interval chosen as momentum space.
  This could have some interesting implications for possible minisuperspace implementation of the GUP theory, 
  concerning the possibility to deal, in the truncated formulation, with wave packets which, differently from the Wheeler-De Witt dynamics \cite{Montani:2009hju}, are slowly spreading even close 
  to the initial singularity or the Big-Bounce, thus allowing, possibly even in that regime, for a quasi-classical approximation of the quantum dynamics. 
  
  We conclude by underlining again that in \cite{Barca:2021epy} and \cite{Battisti:2008am} it has been clarified that the classical GUP dynamics
  (that is a modification of the Poisson brackets in place of the commutators), naturally leads, in the case of an extended formulation with the square root, 
  to the same Friedmann equation for the isotropic Universe, emerging in the Randall-Sundrum model of brane cosmology. 
  Our analysis then opens interesting questions about which of the two proposed extended approaches really corresponds to this singular brane cosmology. 
  As already stated in Section \ref{section_IV}, since the Poisson brackets have been studied without any restrictions on the momentum space, we are led to argue that the above correspondence should be valid for the non-truncated theory. 
  Nevertheless, an intriguing question still would remain on the ground: which kind of cosmological behavior is predicted by the quasi-classical 
  limit of the isotropic Universe in the truncated scenario?

 \appendix*
  \section{}
  
  In ordinary quantum mechanics it is possible to prove that the usual Heisenberg's uncertainty principle (HUP) holds also for those states which are not properly physical states.
  In this brief appendix we want to show how it is possible to extend this result also in the context of the GUP theories defined by the algebra \eqref{generalized_commm1}-\eqref{generalized_commm2}.

 By recalling the definition of physical space, a certain wave function $\Psi$ will be a physical state if it belongs to the domain:
  \begin{equation} \label{physical_space}
 	\mathcal{D}_{\Psi}=\mathcal{D}_{\mathbf{\hat{x}}^2} \cap \mathcal{D}_{\mathbf{\hat{p}}^2} \cap \mathcal{D}_{\mathbf{\hat{x}}\mathbf{\hat{p}}} \cap \mathcal{D}_{\mathbf{\hat{p}}\mathbf{\hat{x}}}.
  \end{equation}
 If this condition is not satisfied, the state $\psi$ fails to be a physical state.
 Yet, for these wave functions it is still possible to define an uncertainty in position and momentum such that the GUP holds.
 Indeed, in the ordinary quantum theory, by use of the Weyl algebra formalism, it is possible to prove that the ordinary Heisenberg's inequality is still valid by taking weaker assumptions on the set to which a generic state $\Psi$ has to belong, in particular it suffices that the state belongs just to the domain of position and momentum operator (\cite{Moretti:2013cma}, section 11.5.6).
 The general validity of this result can be shown to hold also in our case.
 Let consider the operator $\mathbf{\mathbf{\hat{x}}'}= \mathbf{\hat{x}}+a \mathbb{\mathbf{I}}$ and 
 $\mathbf{\mathbf{\hat{p}}'}= \mathbf{\hat{p}}+b\mathbb{\mathbf{I}}$, where $a$ and $b$ are real numbers and $\mathcal{D}_{\mathbf{\mathbf{\hat{x}}'}}= \mathcal{D}_{\mathbf{\hat{x}}}$ and $\mathcal{D}_{\mathbf{\mathbf{\hat{p}}'}}= \mathcal{D}_{\mathbf{\hat{p}}}$.
 
 By explicit computations in p-representation it is easy to show that:
  \begin{equation} \label{exp_val_comm}
  \braket{\mathbf{\mathbf{\hat{x}}'} \psi}{\mathbf{\mathbf{\hat{p}}'} \psi} -  \braket{\mathbf{\mathbf{\hat{p}}'} \psi}{\mathbf{\mathbf{\hat{x}}'} \psi} \!=  \! i \hbar \!\! \int_{\mathbb{R}} \! \! d{p} \abs{\psi(p)}^2, \quad \!\! \forall \psi \! \in \! \mathcal{D}_{\mathbf{\hat{x}}} \cap \mathcal{D}_{\mathbf{\hat{p}}}.
  \end{equation}

 First, we notice that this expression is formally equivalent to the expectation value of the commutator between position and momentum operators for those states for which it can be defined, that is:
   \begin{equation}
	\expval{\comm{\mathbf{\hat{x}}}{\mathbf{\hat{p}}}}= i \hbar \int_{\mathbb{R}} d{p} \abs{\psi(p)}^2, \quad \forall \psi \in \mathcal{D}_{\comm{\mathbf{\hat{x}}}{\mathbf{\hat{p}}}},
   \end{equation}
  but only under the condition:
  \begin{equation}
    \lim_{p \to \pm \infty}   p \psi (p) \partial_p^{(w)}{\psi}^{\ast}(p) = 0.
  \end{equation}

  Then, by choosing $a=\xi$ and $b=\eta$, we can formally write:
  \begin{align} \label{standard_devs}
    \begin{split}
    	\norm{\mathbf{\mathbf{\hat{x}}'}\psi}^2_{\mathcal{L}^2}=\braket{\mathbf{\mathbf{\hat{x}}'}\psi}{\mathbf{\mathbf{\hat{x}}'}\psi} = \int_{\mathbb{R}}  {d\mu_x}(x-\xi)^2  \abs{\psi}^2, & \\
  	    \forall \psi \in \mathcal{D}_{\mathbf{\hat{x}}},& 
  	\end{split}
    \\
    \begin{split}
   	    \norm{\mathbf{\mathbf{\hat{p}}'}\psi}^2_{\mathcal{L}^2}=\braket{\mathbf{\mathbf{\hat{p}}'}\psi}{\mathbf{\mathbf{\hat{p}}'}\psi} = \int_{\mathbb{R}} \frac{dp}{f(p)} (p-\eta)^2  \abs{\psi}^2, & \\
  	    \forall \psi \in \mathcal{D}_{\mathbf{\hat{p}}}, &
  	\end{split}
  \end{align}
  which are the formal expressions of the square standard deviation of the $\mathbf{\hat{x}}$ and $\mathbf{\hat{p}}$ operators in our framework, where with $\mu_x$ we have indicated a generic formal measure in the (possibly non-physical) x-representation. 
  At this point, from \eqref{exp_val_comm} and transposing in the physical p-representation the expression \eqref{standard_devs}, we can infer:
  \begin{equation}
  	\begin{split}
    &\norm{\mathbf{\mathbf{\hat{x}}'}\psi} \norm{\mathbf{\mathbf{\hat{p}}'}\psi} \geq \abs{\braket{\mathbf{\mathbf{\hat{x}}'}\psi}{\mathbf{\mathbf{\hat{p}}'}\psi}}\geq\abs{\Im{\braket{\mathbf{\mathbf{\hat{x}}'}\psi}{\mathbf{\mathbf{\hat{p}}'}\psi}}}\\
    & =\frac{\hbar}{2}\int_{\mathbb{R}} dp \abs{\psi}^2.
    \end{split}
  \end{equation}
  and, in the end, since $\norm{\mathbf{\mathbf{\hat{x}}'}\psi}=\Delta\mathbf{\hat{x}}_{\psi}$ and $\norm{\mathbf{\mathbf{\hat{p}}'}\psi}=\Delta\mathbf{\hat{p}}_{\psi}$:
  \begin{equation}
  	\Delta\mathbf{\hat{x}}_{\psi} \Delta\mathbf{\hat{p}}_{\psi} \geq \frac{\hbar}{2}\int_{\mathbb{R}} dp \abs{\psi}^2=\frac{\hbar}{2}\abs{\expval{f(\mathbf{\hat{p}}}_{\psi}}.
  \end{equation}
  This proves exactly that the GUP is valid for all those states belonging to $\mathcal{D}_{\mathbf{\hat{x}}}\cap \mathcal{D}_{\mathbf{\hat{p}}}$.
  
  It is interesting to notice that this is exactly the case of the maximally localized states of the KMM theory discussed in \cite{Kempf:1994su}.
  The wave function for these states reads as:
  \begin{equation} \label{KMM_max_loc_states}
     \psi_{\xi}^{ml}(p)=\sqrt{\frac{2\sqrt{\beta}}{\pi}}(1+\beta p^2)^{-1}e^{-i\xi \frac{\arctan{(\sqrt{\beta}p})}{\hbar \sqrt{\beta}}},
  \end{equation}
  where $\xi$ is the quasi-position variable.
  
  By inspecting carefully the domains involved in theory, it can be shown that while $\Psi^{ml}_{\xi}$ belongs to 
 $\mathcal{D}_{\mathbf{\hat{x}}^2}$ and $\mathcal{D}_{\mathbf{\hat{x}}\mathbf{\hat{p}}}$, it does not belong to $\mathcal{D}_{\mathbf{\hat{p}}^2}$ and $\mathcal{D}_{\mathbf{\hat{p}}\mathbf{\hat{x}}}$, hence it cannot belong to the whole intersection \footnote{By $\mathbf{\hat{x}}$ operator we are referring to one of the infinite self-adjoint extensions of the operator itself, the domain of which is an extension of $\mathcal{D}_{\mathbf{\hat{x}}}=\mathcal{S}$ and a restriction of $\mathcal{D}_{\mathbf{\hat{x}^\dagger}}$.}.
 
 These considerations lead us to conclude that the functions $\Psi^{ml}_{\xi}$ are not proper physical states.
 
 Nevertheless, on behalf of the result of this appendix, we can conclude that for the maximally localized states \eqref{KMM_max_loc_states} it is still effectively possible to define an uncertainty in position and momentum and a GUP which relates these two quantities, hence validating the whole procedure involving these states at a "kinematic" level, even if, from a "dynamic" perspective, due to the domain to which they belong, these states cannot be considered fully legitimate physical states.

  \bibliography{GUP}

\begin{thebibliography}{41}%
\makeatletter
\providecommand \@ifxundefined [1]{%
 \@ifx{#1\undefined}
}%
\providecommand \@ifnum [1]{%
 \ifnum #1\expandafter \@firstoftwo
 \else \expandafter \@secondoftwo
 \fi
}%
\providecommand \@ifx [1]{%
 \ifx #1\expandafter \@firstoftwo
 \else \expandafter \@secondoftwo
 \fi
}%
\providecommand \natexlab [1]{#1}%
\providecommand \enquote  [1]{``#1''}%
\providecommand \bibnamefont  [1]{#1}%
\providecommand \bibfnamefont [1]{#1}%
\providecommand \citenamefont [1]{#1}%
\providecommand \href@noop [0]{\@secondoftwo}%
\providecommand \href [0]{\begingroup \@sanitize@url \@href}%
\providecommand \@href[1]{\@@startlink{#1}\@@href}%
\providecommand \@@href[1]{\endgroup#1\@@endlink}%
\providecommand \@sanitize@url [0]{\catcode `\\12\catcode `\$12\catcode
  `\&12\catcode `\#12\catcode `\^12\catcode `\_12\catcode `\%12\relax}%
\providecommand \@@startlink[1]{}%
\providecommand \@@endlink[0]{}%
\providecommand \url  [0]{\begingroup\@sanitize@url \@url }%
\providecommand \@url [1]{\endgroup\@href {#1}{\urlprefix }}%
\providecommand \urlprefix  [0]{URL }%
\providecommand \Eprint [0]{\href }%
\providecommand \doibase [0]{http://dx.doi.org/}%
\providecommand \selectlanguage [0]{\@gobble}%
\providecommand \bibinfo  [0]{\@secondoftwo}%
\providecommand \bibfield  [0]{\@secondoftwo}%
\providecommand \translation [1]{[#1]}%
\providecommand \BibitemOpen [0]{}%
\providecommand \bibitemStop [0]{}%
\providecommand \bibitemNoStop [0]{.\EOS\space}%
\providecommand \EOS [0]{\spacefactor3000\relax}%
\providecommand \BibitemShut  [1]{\csname bibitem#1\endcsname}%
\let\auto@bib@innerbib\@empty
\bibitem [{\citenamefont {Cianfrani}\ \emph {et~al.}(2014)\citenamefont
  {Cianfrani}, \citenamefont {Lecian}, \citenamefont {Lulli},\ and\
  \citenamefont {Montani}}]{CQG:2014}%
  \BibitemOpen
  \bibfield  {author} {\bibinfo {author} {\bibfnamefont {F.}~\bibnamefont
  {Cianfrani}}, \bibinfo {author} {\bibfnamefont {O.~M.}\ \bibnamefont
  {Lecian}}, \bibinfo {author} {\bibfnamefont {M.}~\bibnamefont {Lulli}}, \
  and\ \bibinfo {author} {\bibfnamefont {G.}~\bibnamefont {Montani}},\ }\href
  {\doibase 10.1142/8957} {\emph {\bibinfo {title} {Canonical Quantum Gravity:
  Fundamentals and Recent Developments}}}\ (\bibinfo  {publisher} {World
  Scientific},\ \bibinfo {address} {Singapore},\ \bibinfo {year}
  {2014})\BibitemShut {NoStop}%
\bibitem [{\citenamefont {Rovelli}(2004)}]{Rovelli:2004tv}%
  \BibitemOpen
  \bibfield  {author} {\bibinfo {author} {\bibfnamefont {C.}~\bibnamefont
  {Rovelli}},\ }\href {\doibase 10.1017/CBO9780511755804} {\emph {\bibinfo
  {title} {Quantum gravity}}},\ Cambridge Monographs on Mathematical Physics\
  (\bibinfo  {publisher} {Univ. Pr.},\ \bibinfo {address} {Cambridge, UK},\
  \bibinfo {year} {2004})\BibitemShut {NoStop}%
\bibitem [{\citenamefont {Polchinski}(2017)}]{Polchinski:2014mva}%
  \BibitemOpen
  \bibfield  {author} {\bibinfo {author} {\bibfnamefont {J.}~\bibnamefont
  {Polchinski}},\ }\href {\doibase 10.1016/j.shpsb.2015.08.011} {\bibfield
  {journal} {\bibinfo  {journal} {Stud. Hist. Phil. Sci. B}\ }\textbf {\bibinfo
  {volume} {59}},\ \bibinfo {pages} {6} (\bibinfo {year} {2017})},\ \Eprint
  {http://arxiv.org/abs/1412.5704}{arXiv:1412.5704 [hep-th]}\BibitemShut
  {NoStop}%
\bibitem [{\citenamefont {Amati}\ \emph {et~al.}(1987)\citenamefont {Amati},
  \citenamefont {Ciafaloni},\ and\ \citenamefont {Veneziano}}]{Amati:1987wq}%
  \BibitemOpen
  \bibfield  {author} {\bibinfo {author} {\bibfnamefont {D.}~\bibnamefont
  {Amati}}, \bibinfo {author} {\bibfnamefont {M.}~\bibnamefont {Ciafaloni}}, \
  and\ \bibinfo {author} {\bibfnamefont {G.}~\bibnamefont {Veneziano}},\ }\href
  {\doibase 10.1016/0370-2693(87)90346-7} {\bibfield  {journal} {\bibinfo
  {journal} {Phys. Lett. B}\ }\textbf {\bibinfo {volume} {197}},\ \bibinfo
  {pages} {81} (\bibinfo {year} {1987})}\BibitemShut {NoStop}%
\bibitem [{\citenamefont {Amati}\ \emph {et~al.}(1989)\citenamefont {Amati},
  \citenamefont {Ciafaloni},\ and\ \citenamefont {Veneziano}}]{Amati:1988tn}%
  \BibitemOpen
  \bibfield  {author} {\bibinfo {author} {\bibfnamefont {D.}~\bibnamefont
  {Amati}}, \bibinfo {author} {\bibfnamefont {M.}~\bibnamefont {Ciafaloni}}, \
  and\ \bibinfo {author} {\bibfnamefont {G.}~\bibnamefont {Veneziano}},\ }\href
  {\doibase 10.1016/0370-2693(89)91366-X} {\bibfield  {journal} {\bibinfo
  {journal} {Phys. Lett. B}\ }\textbf {\bibinfo {volume} {216}},\ \bibinfo
  {pages} {41} (\bibinfo {year} {1989})}\BibitemShut {NoStop}%
\bibitem [{\citenamefont {Gross}\ and\ \citenamefont
  {Mende}(1988)}]{Gross:1987ar}%
  \BibitemOpen
  \bibfield  {author} {\bibinfo {author} {\bibfnamefont {D.~J.}\ \bibnamefont
  {Gross}}\ and\ \bibinfo {author} {\bibfnamefont {P.~F.}\ \bibnamefont
  {Mende}},\ }\href {\doibase 10.1016/0550-3213(88)90390-2} {\bibfield
  {journal} {\bibinfo  {journal} {Nucl. Phys. B}\ }\textbf {\bibinfo {volume}
  {303}},\ \bibinfo {pages} {407} (\bibinfo {year} {1988})}\BibitemShut
  {NoStop}%
\bibitem [{\citenamefont {Gross}\ and\ \citenamefont
  {Mende}(1987)}]{Gross:1987kza}%
  \BibitemOpen
  \bibfield  {author} {\bibinfo {author} {\bibfnamefont {D.~J.}\ \bibnamefont
  {Gross}}\ and\ \bibinfo {author} {\bibfnamefont {P.~F.}\ \bibnamefont
  {Mende}},\ }\href {\doibase 10.1016/0370-2693(87)90355-8} {\bibfield
  {journal} {\bibinfo  {journal} {Phys. Lett. B}\ }\textbf {\bibinfo {volume}
  {197}},\ \bibinfo {pages} {129} (\bibinfo {year} {1987})}\BibitemShut
  {NoStop}%
\bibitem [{\citenamefont {Konishi}\ \emph {et~al.}(1990)\citenamefont
  {Konishi}, \citenamefont {Paffuti},\ and\ \citenamefont
  {Provero}}]{Konishi:1989wk}%
  \BibitemOpen
  \bibfield  {author} {\bibinfo {author} {\bibfnamefont {K.}~\bibnamefont
  {Konishi}}, \bibinfo {author} {\bibfnamefont {G.}~\bibnamefont {Paffuti}}, \
  and\ \bibinfo {author} {\bibfnamefont {P.}~\bibnamefont {Provero}},\ }\href
  {\doibase 10.1016/0370-2693(90)91927-4} {\bibfield  {journal} {\bibinfo
  {journal} {Phys. Lett. B}\ }\textbf {\bibinfo {volume} {234}},\ \bibinfo
  {pages} {276} (\bibinfo {year} {1990})}\BibitemShut {NoStop}%
\bibitem [{\citenamefont {Kempf}(1994{\natexlab{a}})}]{Kempf:1993bq}%
  \BibitemOpen
  \bibfield  {author} {\bibinfo {author} {\bibfnamefont {A.}~\bibnamefont
  {Kempf}},\ }\href {\doibase 10.1063/1.530798} {\bibfield  {journal} {\bibinfo
   {journal} {J. Math. Phys.}\ }\textbf {\bibinfo {volume} {35}},\ \bibinfo
  {pages} {4483} (\bibinfo {year} {1994}{\natexlab{a}})},\ \Eprint
  {http://arxiv.org/abs/hep-th/9311147}{arXiv:hep-th/9311147}\BibitemShut
  {NoStop}%
\bibitem [{\citenamefont {Kempf}(1994{\natexlab{b}})}]{Kempf:1994qp}%
  \BibitemOpen
  \bibfield  {author} {\bibinfo {author} {\bibfnamefont {A.}~\bibnamefont
  {Kempf}},\ }\href@noop {} {\  (\bibinfo {year} {1994}{\natexlab{b}})},\
  \Eprint
  {http://arxiv.org/abs/hep-th/9405067}{arXiv:hep-th/9405067}\BibitemShut
  {NoStop}%
\bibitem [{\citenamefont {Kempf}\ \emph {et~al.}(1995)\citenamefont {Kempf},
  \citenamefont {Mangano},\ and\ \citenamefont {Mann}}]{Kempf:1994su}%
  \BibitemOpen
  \bibfield  {author} {\bibinfo {author} {\bibfnamefont {A.}~\bibnamefont
  {Kempf}}, \bibinfo {author} {\bibfnamefont {G.}~\bibnamefont {Mangano}}, \
  and\ \bibinfo {author} {\bibfnamefont {R.~B.}\ \bibnamefont {Mann}},\ }\href
  {\doibase 10.1103/PhysRevD.52.1108} {\bibfield  {journal} {\bibinfo
  {journal} {Phys. Rev. D}\ }\textbf {\bibinfo {volume} {52}},\ \bibinfo
  {pages} {1108} (\bibinfo {year} {1995})},\ \Eprint
  {http://arxiv.org/abs/hep-th/9412167}{arXiv:hep-th/9412167}\BibitemShut
  {NoStop}%
\bibitem [{\citenamefont {Ashtekar}\ and\ \citenamefont
  {Singh}(2011)}]{Ashtekar:2011ni}%
  \BibitemOpen
  \bibfield  {author} {\bibinfo {author} {\bibfnamefont {A.}~\bibnamefont
  {Ashtekar}}\ and\ \bibinfo {author} {\bibfnamefont {P.}~\bibnamefont
  {Singh}},\ }\href {\doibase 10.1088/0264-9381/28/21/213001} {\bibfield
  {journal} {\bibinfo  {journal} {Class. Quant. Grav.}\ }\textbf {\bibinfo
  {volume} {28}},\ \bibinfo {pages} {213001} (\bibinfo {year} {2011})},\
  \Eprint {http://arxiv.org/abs/1108.0893}{arXiv:1108.0893 [gr-qc]}\BibitemShut
  {NoStop}%
\bibitem [{\citenamefont {Corichi}\ \emph {et~al.}(2007)\citenamefont
  {Corichi}, \citenamefont {Vukasinac},\ and\ \citenamefont
  {Zapata}}]{Corichi:2007tf}%
  \BibitemOpen
  \bibfield  {author} {\bibinfo {author} {\bibfnamefont {A.}~\bibnamefont
  {Corichi}}, \bibinfo {author} {\bibfnamefont {T.}~\bibnamefont {Vukasinac}},
  \ and\ \bibinfo {author} {\bibfnamefont {J.~A.}\ \bibnamefont {Zapata}},\
  }\href {\doibase 10.1103/PhysRevD.76.044016} {\bibfield  {journal} {\bibinfo
  {journal} {Phys. Rev. D}\ }\textbf {\bibinfo {volume} {76}},\ \bibinfo
  {pages} {044016} (\bibinfo {year} {2007})},\ \Eprint
  {http://arxiv.org/abs/0704.0007}{arXiv:0704.0007 [gr-qc]}\BibitemShut
  {NoStop}%
\bibitem [{\citenamefont {Barca}\ \emph {et~al.}(2021)\citenamefont {Barca},
  \citenamefont {Giovannetti},\ and\ \citenamefont {Montani}}]{Barca:2021qdn}%
  \BibitemOpen
  \bibfield  {author} {\bibinfo {author} {\bibfnamefont {G.}~\bibnamefont
  {Barca}}, \bibinfo {author} {\bibfnamefont {E.}~\bibnamefont {Giovannetti}},
  \ and\ \bibinfo {author} {\bibfnamefont {G.}~\bibnamefont {Montani}},\ }\href
  {\doibase 10.3390/universe7090327} {\bibfield  {journal} {\bibinfo  {journal}
  {Universe}\ }\textbf {\bibinfo {volume} {7}},\ \bibinfo {pages} {327}
  (\bibinfo {year} {2021})},\ \Eprint
  {http://arxiv.org/abs/2109.08645}{arXiv:2109.08645 [gr-qc]}\BibitemShut
  {NoStop}%
\bibitem [{\citenamefont {Barca}\ \emph {et~al.}(2022)\citenamefont {Barca},
  \citenamefont {Giovannetti},\ and\ \citenamefont {Montani}}]{Barca:2021epy}%
  \BibitemOpen
  \bibfield  {author} {\bibinfo {author} {\bibfnamefont {G.}~\bibnamefont
  {Barca}}, \bibinfo {author} {\bibfnamefont {E.}~\bibnamefont {Giovannetti}},
  \ and\ \bibinfo {author} {\bibfnamefont {G.}~\bibnamefont {Montani}},\ }\href
  {\doibase 10.1142/S0219887822500979} {\bibfield  {journal} {\bibinfo
  {journal} {Int. J. Geom. Meth. Mod. Phys.}\ }\textbf {\bibinfo {volume}
  {19}},\ \bibinfo {pages} {2250097} (\bibinfo {year} {2022})},\ \Eprint
  {http://arxiv.org/abs/2112.08905}{arXiv:2112.08905 [gr-qc]}\BibitemShut
  {NoStop}%
\bibitem [{\citenamefont {Bosso}(2021)}]{Bosso:2020aqm}%
  \BibitemOpen
  \bibfield  {author} {\bibinfo {author} {\bibfnamefont {P.}~\bibnamefont
  {Bosso}},\ }\href {\doibase 10.1088/1361-6382/abe758} {\bibfield  {journal}
  {\bibinfo  {journal} {Class. Quant. Grav.}\ }\textbf {\bibinfo {volume}
  {38}},\ \bibinfo {pages} {075021} (\bibinfo {year} {2021})},\ \Eprint
  {http://arxiv.org/abs/2005.12258}{arXiv:2005.12258 [gr-qc]}\BibitemShut
  {NoStop}%
\bibitem [{\citenamefont {Bosso}\ and\ \citenamefont
  {Luciano}(2021)}]{Bosso:2021koi}%
  \BibitemOpen
  \bibfield  {author} {\bibinfo {author} {\bibfnamefont {P.}~\bibnamefont
  {Bosso}}\ and\ \bibinfo {author} {\bibfnamefont {G.~G.}\ \bibnamefont
  {Luciano}},\ }\href {\doibase 10.1140/epjc/s10052-021-09795-1} {\bibfield
  {journal} {\bibinfo  {journal} {Eur. Phys. J. C}\ }\textbf {\bibinfo {volume}
  {81}},\ \bibinfo {pages} {982} (\bibinfo {year} {2021})},\ \Eprint
  {http://arxiv.org/abs/2109.15259}{arXiv:2109.15259 [hep-th]}\BibitemShut
  {NoStop}%
\bibitem [{\citenamefont {Bosso}(2022)}]{Bosso:2022rue}%
  \BibitemOpen
  \bibfield  {author} {\bibinfo {author} {\bibfnamefont {P.}~\bibnamefont
  {Bosso}},\ }\href@noop {} {\  (\bibinfo {year} {2022})},\ \Eprint
  {http://arxiv.org/abs/2206.15422}{arXiv:2206.15422 [gr-qc]}\BibitemShut
  {NoStop}%
\bibitem [{\citenamefont {Bosso}\ \emph {et~al.}(2022)\citenamefont {Bosso},
  \citenamefont {Petruzziello},\ and\ \citenamefont {Wagner}}]{Bosso:2022vlz}%
  \BibitemOpen
  \bibfield  {author} {\bibinfo {author} {\bibfnamefont {P.}~\bibnamefont
  {Bosso}}, \bibinfo {author} {\bibfnamefont {L.}~\bibnamefont {Petruzziello}},
  \ and\ \bibinfo {author} {\bibfnamefont {F.}~\bibnamefont {Wagner}},\ }\href
  {\doibase 10.1016/j.physletb.2022.137415} {\bibfield  {journal} {\bibinfo
  {journal} {Phys. Lett. B}\ }\textbf {\bibinfo {volume} {834}},\ \bibinfo
  {pages} {137415} (\bibinfo {year} {2022})},\ \Eprint
  {http://arxiv.org/abs/2206.05064}{arXiv:2206.05064 [gr-qc]}\BibitemShut
  {NoStop}%
\bibitem [{\citenamefont {Battisti}\ and\ \citenamefont
  {Montani}(2007)}]{Battisti:2007jd}%
  \BibitemOpen
  \bibfield  {author} {\bibinfo {author} {\bibfnamefont {M.~V.}\ \bibnamefont
  {Battisti}}\ and\ \bibinfo {author} {\bibfnamefont {G.}~\bibnamefont
  {Montani}},\ }\href {\doibase 10.1016/j.physletb.2007.09.012} {\bibfield
  {journal} {\bibinfo  {journal} {Phys. Lett. B}\ }\textbf {\bibinfo {volume}
  {656}},\ \bibinfo {pages} {96} (\bibinfo {year} {2007})},\ \Eprint
  {http://arxiv.org/abs/gr-qc/0703025}{arXiv:gr-qc/0703025}\BibitemShut
  {NoStop}%
\bibitem [{\citenamefont {Battisti}\ and\ \citenamefont
  {Montani}(2008{\natexlab{a}})}]{Battisti:2007zg}%
  \BibitemOpen
  \bibfield  {author} {\bibinfo {author} {\bibfnamefont {M.~V.}\ \bibnamefont
  {Battisti}}\ and\ \bibinfo {author} {\bibfnamefont {G.}~\bibnamefont
  {Montani}},\ }\href {\doibase 10.1103/PhysRevD.77.023518} {\bibfield
  {journal} {\bibinfo  {journal} {Phys. Rev. D}\ }\textbf {\bibinfo {volume}
  {77}},\ \bibinfo {pages} {023518} (\bibinfo {year} {2008}{\natexlab{a}})},\
  \Eprint {http://arxiv.org/abs/0707.2726}{arXiv:0707.2726 [gr-qc]}\BibitemShut
  {NoStop}%
\bibitem [{\citenamefont {Battisti}\ and\ \citenamefont
  {Montani}(2008{\natexlab{b}})}]{Battisti:2008rv}%
  \BibitemOpen
  \bibfield  {author} {\bibinfo {author} {\bibfnamefont {M.~V.}\ \bibnamefont
  {Battisti}}\ and\ \bibinfo {author} {\bibfnamefont {G.}~\bibnamefont
  {Montani}},\ }\href {\doibase 10.1142/S0217751X08040184} {\bibfield
  {journal} {\bibinfo  {journal} {Int. J. Mod. Phys. A}\ }\textbf {\bibinfo
  {volume} {23}},\ \bibinfo {pages} {1257} (\bibinfo {year}
  {2008}{\natexlab{b}})},\ \Eprint
  {http://arxiv.org/abs/0802.0688}{arXiv:0802.0688 [gr-qc]}\BibitemShut
  {NoStop}%
\bibitem [{\citenamefont {Battisti}\ and\ \citenamefont
  {Montani}(2009{\natexlab{a}})}]{Battisti:2008qi}%
  \BibitemOpen
  \bibfield  {author} {\bibinfo {author} {\bibfnamefont {M.~V.}\ \bibnamefont
  {Battisti}}\ and\ \bibinfo {author} {\bibfnamefont {G.}~\bibnamefont
  {Montani}},\ }\href {\doibase 10.1016/j.physletb.2009.10.003} {\bibfield
  {journal} {\bibinfo  {journal} {Phys. Lett. B}\ }\textbf {\bibinfo {volume}
  {681}},\ \bibinfo {pages} {179} (\bibinfo {year} {2009}{\natexlab{a}})},\
  \Eprint {http://arxiv.org/abs/0808.0831}{arXiv:0808.0831 [gr-qc]}\BibitemShut
  {NoStop}%
\bibitem [{\citenamefont {Battisti}(2009)}]{Battisti:2008am}%
  \BibitemOpen
  \bibfield  {author} {\bibinfo {author} {\bibfnamefont {M.~V.}\ \bibnamefont
  {Battisti}},\ }\href {\doibase 10.1088/1742-6596/189/1/012005} {\bibfield
  {journal} {\bibinfo  {journal} {J. Phys. Conf. Ser.}\ }\textbf {\bibinfo
  {volume} {189}},\ \bibinfo {pages} {012005} (\bibinfo {year} {2009})},\
  \Eprint {http://arxiv.org/abs/0810.5039}{arXiv:0810.5039 [gr-qc]}\BibitemShut
  {NoStop}%
\bibitem [{\citenamefont {Battisti}\ and\ \citenamefont
  {Montani}(2009{\natexlab{b}})}]{Battisti:2009at}%
  \BibitemOpen
  \bibfield  {author} {\bibinfo {author} {\bibfnamefont {M.~V.}\ \bibnamefont
  {Battisti}}\ and\ \bibinfo {author} {\bibfnamefont {G.}~\bibnamefont
  {Montani}},\ }\bibfield  {booktitle} {\emph {\bibinfo {booktitle} {3rd
  Steckelberg Workshop on Relativistic Field Theories}},\ }\href@noop {} {\
  (\bibinfo {year} {2009}{\natexlab{b}})},\ \Eprint
  {http://arxiv.org/abs/0903.0494}{arXiv:0903.0494 [gr-qc]}\BibitemShut
  {NoStop}%
\bibitem [{\citenamefont {Nowakowski}\ and\ \citenamefont
  {Arraut}(2009)}]{Nowakowski:2009ha}%
  \BibitemOpen
  \bibfield  {author} {\bibinfo {author} {\bibfnamefont {M.}~\bibnamefont
  {Nowakowski}}\ and\ \bibinfo {author} {\bibfnamefont {I.}~\bibnamefont
  {Arraut}},\ }\href {\doibase 10.1142/S0217732309030679} {\bibfield  {journal}
  {\bibinfo  {journal} {Mod. Phys. Lett. A}\ }\textbf {\bibinfo {volume}
  {24}},\ \bibinfo {pages} {2133} (\bibinfo {year} {2009})},\ \Eprint
  {http://arxiv.org/abs/0905.3762}{arXiv:0905.3762 [gr-qc]}\BibitemShut
  {NoStop}%
\bibitem [{\citenamefont {Arraut}\ \emph {et~al.}(2009)\citenamefont {Arraut},
  \citenamefont {Batic},\ and\ \citenamefont {Nowakowski}}]{Arraut:2008hc}%
  \BibitemOpen
  \bibfield  {author} {\bibinfo {author} {\bibfnamefont {I.}~\bibnamefont
  {Arraut}}, \bibinfo {author} {\bibfnamefont {D.}~\bibnamefont {Batic}}, \
  and\ \bibinfo {author} {\bibfnamefont {M.}~\bibnamefont {Nowakowski}},\
  }\href {\doibase 10.1088/0264-9381/26/12/125006} {\bibfield  {journal}
  {\bibinfo  {journal} {Class. Quant. Grav.}\ }\textbf {\bibinfo {volume}
  {26}},\ \bibinfo {pages} {125006} (\bibinfo {year} {2009})},\ \Eprint
  {http://arxiv.org/abs/0810.5156}{arXiv:0810.5156 [gr-qc]}\BibitemShut
  {NoStop}%
\bibitem [{\citenamefont {Arraut}(2012)}]{Arraut:2012yd}%
  \BibitemOpen
  \bibfield  {author} {\bibinfo {author} {\bibfnamefont {I.}~\bibnamefont
  {Arraut}},\ }\href@noop {} {\  (\bibinfo {year} {2012})},\ \Eprint
  {http://arxiv.org/abs/1205.6905}{arXiv:1205.6905 [gr-qc]}\BibitemShut
  {NoStop}%
\bibitem [{\citenamefont {Fadel}\ and\ \citenamefont
  {Maggiore}(2022)}]{Fadel:2021hnx}%
  \BibitemOpen
  \bibfield  {author} {\bibinfo {author} {\bibfnamefont {M.}~\bibnamefont
  {Fadel}}\ and\ \bibinfo {author} {\bibfnamefont {M.}~\bibnamefont
  {Maggiore}},\ }\href {\doibase 10.1103/PhysRevD.105.106017} {\bibfield
  {journal} {\bibinfo  {journal} {Phys. Rev. D}\ }\textbf {\bibinfo {volume}
  {105}},\ \bibinfo {pages} {106017} (\bibinfo {year} {2022})},\ \Eprint
  {http://arxiv.org/abs/2112.09034}{arXiv:2112.09034 [quant-ph]}\BibitemShut
  {NoStop}%
\bibitem [{\citenamefont {Maggiore}(1993{\natexlab{a}})}]{Maggiore:1993rv}%
  \BibitemOpen
  \bibfield  {author} {\bibinfo {author} {\bibfnamefont {M.}~\bibnamefont
  {Maggiore}},\ }\href {\doibase 10.1016/0370-2693(93)91401-8} {\bibfield
  {journal} {\bibinfo  {journal} {Phys. Lett. B}\ }\textbf {\bibinfo {volume}
  {304}},\ \bibinfo {pages} {65} (\bibinfo {year} {1993}{\natexlab{a}})},\
  \Eprint
  {http://arxiv.org/abs/hep-th/9301067}{arXiv:hep-th/9301067}\BibitemShut
  {NoStop}%
\bibitem [{\citenamefont {Maggiore}(1993{\natexlab{b}})}]{Maggiore:1993kv}%
  \BibitemOpen
  \bibfield  {author} {\bibinfo {author} {\bibfnamefont {M.}~\bibnamefont
  {Maggiore}},\ }\href {\doibase 10.1016/0370-2693(93)90785-G} {\bibfield
  {journal} {\bibinfo  {journal} {Phys. Lett. B}\ }\textbf {\bibinfo {volume}
  {319}},\ \bibinfo {pages} {83} (\bibinfo {year} {1993}{\natexlab{b}})},\
  \Eprint
  {http://arxiv.org/abs/hep-th/9309034}{arXiv:hep-th/9309034}\BibitemShut
  {NoStop}%
\bibitem [{\citenamefont {Maggiore}(1994)}]{Maggiore:1993zu}%
  \BibitemOpen
  \bibfield  {author} {\bibinfo {author} {\bibfnamefont {M.}~\bibnamefont
  {Maggiore}},\ }\href {\doibase 10.1103/PhysRevD.49.5182} {\bibfield
  {journal} {\bibinfo  {journal} {Phys. Rev. D}\ }\textbf {\bibinfo {volume}
  {49}},\ \bibinfo {pages} {5182} (\bibinfo {year} {1994})},\ \Eprint
  {http://arxiv.org/abs/hep-th/9305163}{arXiv:hep-th/9305163}\BibitemShut
  {NoStop}%
\bibitem [{\citenamefont {Randall}\ and\ \citenamefont
  {Sundrum}(1999)}]{Randall:1999vf}%
  \BibitemOpen
  \bibfield  {author} {\bibinfo {author} {\bibfnamefont {L.}~\bibnamefont
  {Randall}}\ and\ \bibinfo {author} {\bibfnamefont {R.}~\bibnamefont
  {Sundrum}},\ }\href {\doibase 10.1103/PhysRevLett.83.4690} {\bibfield
  {journal} {\bibinfo  {journal} {Phys. Rev. Lett.}\ }\textbf {\bibinfo
  {volume} {83}},\ \bibinfo {pages} {4690} (\bibinfo {year} {1999})},\ \Eprint
  {http://arxiv.org/abs/hep-th/9906064}{arXiv:hep-th/9906064}\BibitemShut
  {NoStop}%
\bibitem [{\citenamefont {Ashtekar}\ and\ \citenamefont
  {Gupt}(2015)}]{Ashtekar:2015iza}%
  \BibitemOpen
  \bibfield  {author} {\bibinfo {author} {\bibfnamefont {A.}~\bibnamefont
  {Ashtekar}}\ and\ \bibinfo {author} {\bibfnamefont {B.}~\bibnamefont
  {Gupt}},\ }\href {\doibase 10.1103/PhysRevD.92.084060} {\bibfield  {journal}
  {\bibinfo  {journal} {Phys. Rev. D}\ }\textbf {\bibinfo {volume} {92}},\
  \bibinfo {pages} {084060} (\bibinfo {year} {2015})},\ \Eprint
  {http://arxiv.org/abs/1509.08899}{arXiv:1509.08899 [gr-qc]}\BibitemShut
  {NoStop}%
\bibitem [{\citenamefont {Gomes}(2022)}]{Gomes:2022}%
  \BibitemOpen
  \bibfield  {author} {\bibinfo {author} {\bibfnamefont {A.~H.}\ \bibnamefont
  {Gomes}},\ }\href@noop {} {\  (\bibinfo {year} {2022})},\ \Eprint
  {http://arxiv.org/abs/2202.02044}{arXiv:2202.02044 [quant-ph]}\BibitemShut
  {NoStop}%
\bibitem [{\citenamefont {Liebmann}\ \emph {et~al.}(2019)\citenamefont
  {Liebmann}, \citenamefont {R{\"u}haak},\ and\ \citenamefont
  {Henschenmacher}}]{liebmann2019non}%
  \BibitemOpen
  \bibfield  {author} {\bibinfo {author} {\bibfnamefont {M.}~\bibnamefont
  {Liebmann}}, \bibinfo {author} {\bibfnamefont {H.}~\bibnamefont
  {R{\"u}haak}}, \ and\ \bibinfo {author} {\bibfnamefont {B.}~\bibnamefont
  {Henschenmacher}},\ }\href@noop {} {\  (\bibinfo {year} {2019})},\ \Eprint
  {http://arxiv.org/abs/1909.04027}{arXiv:1909.04027}\BibitemShut {NoStop}%
\bibitem [{\citenamefont {Dzhunushaliev}(2006)}]{Dzhunushaliev:2005yd}%
  \BibitemOpen
  \bibfield  {author} {\bibinfo {author} {\bibfnamefont {V.}~\bibnamefont
  {Dzhunushaliev}},\ }\href {\doibase 10.1007/s10702-006-0373-2} {\bibfield
  {journal} {\bibinfo  {journal} {Found. Phys. Lett.}\ }\textbf {\bibinfo
  {volume} {19}},\ \bibinfo {pages} {157} (\bibinfo {year} {2006})},\ \Eprint
  {http://arxiv.org/abs/hep-th/0502216}{arXiv:hep-th/0502216}\BibitemShut
  {NoStop}%
\bibitem [{\citenamefont {Detournay}\ \emph {et~al.}(2002)\citenamefont
  {Detournay}, \citenamefont {Gabriel},\ and\ \citenamefont
  {Spindel}}]{Detournay:2002fq}%
  \BibitemOpen
  \bibfield  {author} {\bibinfo {author} {\bibfnamefont {S.}~\bibnamefont
  {Detournay}}, \bibinfo {author} {\bibfnamefont {C.}~\bibnamefont {Gabriel}},
  \ and\ \bibinfo {author} {\bibfnamefont {P.}~\bibnamefont {Spindel}},\ }\href
  {\doibase 10.1103/PhysRevD.66.125004} {\bibfield  {journal} {\bibinfo
  {journal} {Phys. Rev. D}\ }\textbf {\bibinfo {volume} {66}},\ \bibinfo
  {pages} {125004} (\bibinfo {year} {2002})},\ \Eprint
  {http://arxiv.org/abs/hep-th/0210128}{arXiv:hep-th/0210128}\BibitemShut
  {NoStop}%
\bibitem [{\citenamefont {Guendelman}(2021)}]{Guendelman:2021zmf}%
  \BibitemOpen
  \bibfield  {author} {\bibinfo {author} {\bibfnamefont {E.~I.}\ \bibnamefont
  {Guendelman}},\ }\href {\doibase 10.1142/S0218271821420281} {\bibfield
  {journal} {\bibinfo  {journal} {Int. J. Mod. Phys. D}\ }\textbf {\bibinfo
  {volume} {30}},\ \bibinfo {pages} {2142028} (\bibinfo {year} {2021})},\
  \Eprint {http://arxiv.org/abs/2110.09199}{arXiv:2110.09199
  [hep-th]}\BibitemShut {NoStop}%
\bibitem [{\citenamefont {Montani}\ \emph {et~al.}(2009)\citenamefont
  {Montani}, \citenamefont {Battisti}, \citenamefont {Benini},\ and\
  \citenamefont {Imponente}}]{Montani:2009hju}%
  \BibitemOpen
  \bibfield  {author} {\bibinfo {author} {\bibfnamefont {G.}~\bibnamefont
  {Montani}}, \bibinfo {author} {\bibfnamefont {M.~V.}\ \bibnamefont
  {Battisti}}, \bibinfo {author} {\bibfnamefont {R.}~\bibnamefont {Benini}}, \
  and\ \bibinfo {author} {\bibfnamefont {G.}~\bibnamefont {Imponente}},\ }\href
  {\doibase https://doi.org/10.1142/7235} {\emph {\bibinfo {title} {Primordial
  cosmology}}}\ (\bibinfo  {publisher} {World Scientific},\ \bibinfo {address}
  {Singapore},\ \bibinfo {year} {2009})\BibitemShut {NoStop}%
\bibitem [{\citenamefont {Moretti}(2013)}]{Moretti:2013cma}%
  \BibitemOpen
  \bibfield  {author} {\bibinfo {author} {\bibfnamefont {V.}~\bibnamefont
  {Moretti}},\ }\href {\doibase 10.1007/978-3-319-70706-8} {\emph {\bibinfo
  {title} {Spectral Theory and Quantum Mechanics: Mathematical Foundations of
  Quantum Theories, Symmetries and Introduction to the Algebraic
  Formulation}}},\ \bibinfo {edition} {2nd}\ ed.,\ \bibinfo {series} {UNITEXT:
  La Matematica per il 3+2}, Vol.\ \bibinfo {volume} {110}\ (\bibinfo
  {publisher} {Springer},\ \bibinfo {year} {2013})\BibitemShut {NoStop}%
\end{thebibliography}%
  \bibliographystyle{apsrev4-2}
 \end{document}